\documentclass[twocolumn,tighten,trackchanges]{aastex701}

\usepackage{soul}

\newcommand{\del}[1]{\textcolor{magenta}{{\iffalse{#1}\fi}}}

\usepackage{tikz}
\usepackage{amsmath}

\begin{document}

\title{The PSR J0435+3233 Triple System}


\author[orcid=0009-0009-6590-1540]{Z.~L. Yang}
\affiliation{National Astronomical Observatories, Chinese Academy of Sciences, Jia-20 Datun Road, ChaoYang District, Beijing 100012, China}
\email[notshow]{zlyang@nao.cas.cn}  

\author[orcid=0000-0002-9274-3092,gname=JinLin,sname=Han]{J.~L. Han}\thanks{E-mail: hjl@nao.cas.cn}
\affiliation{National Astronomical Observatories, Chinese Academy of Sciences, Jia-20 Datun Road, ChaoYang District, Beijing 100012, China}
\affiliation{School of Astronomy and Space Science, University of Chinese Academy of Sciences, Beijing 100049, China}
\affiliation{State Key Laboratory of Radio Astronomy and Technology, Beijing 100101, China }
\email[notshow]{hjl@nao.cas.cn}  

\author[0009-0008-1612-9948]{Yi Yan}
\affiliation{National Astronomical Observatories, Chinese Academy of Sciences, Jia-20 Datun Road, ChaoYang District, Beijing 100012, China}
\email[notshow]{yanyi@nao.cas.cn}  

\author[0000-0002-0643-8295]{Bin Liu}
\affiliation{Institute for Astronomy, School of Physics, Zhejiang University, 310058 Hangzhou, China}
\email[notshow]{liubin23@zju.edu.cn}  

\author[0009-0008-2437-0184]{Y.~L. Guo}
\affiliation{International Center of Supernovae (ICESUN), Yunnan Key Laboratory of Supernova Research, Yunnan Observatories, Chinese Academy of Sciences (CAS), Kunming 650216, China}
\email[notshow]{yunlang@ynao.ac.cn}  

\author[0009-0000-6595-2537]{M.K. Yang}
\affiliation{International Centre of Supernovae (ICESUN), Yunnan Key Laboratory of Supernova Research, Yunnan Observatories, Chinese Academy of Sciences (CAS), Kunming 650216, China}
\email[notshow]{yangmk@bao.ac.cn}  

\author[0000-0002-3231-1167]{Bo Wang}
\affiliation{International Center of Supernovae (ICESUN), Yunnan Key Laboratory of Supernova Research, Yunnan Observatories, Chinese Academy of Sciences (CAS), Kunming 650216, China}
\email[notshow]{wangbo@ynao.ac.cn}  

\author[0000-0003-3137-1851]{Wei-Min Gu}
\affiliation{Department of Astronomy, Xiamen University, Xiamen, Fujian 361005, People's Republic of China}
\email[notshow]{guwm@xmu.edu.cn}  

\author[0000-0002-2577-1990]{J. Li}
\affiliation{International Center of Supernovae (ICESUN), Yunnan Key Laboratory of Supernova Research, Yunnan Observatories, Chinese Academy of Sciences (CAS), Kunming 650216, China}
\email[notshow]{lijiao@ynao.ac.cn}  

\author[0009-0007-5354-2611]{L. H. Li}
\affiliation{International Center of Supernovae (ICESUN), Yunnan Key Laboratory of Supernova Research, Yunnan Observatories, Chinese Academy of Sciences (CAS), Kunming 650216, China}
\email[notshow]{liluhan@ynan.ac.cn}  

\author[0000-0003-1778-5580]{J. Xu}
\affiliation{National Astronomical Observatories, Chinese Academy of Sciences, Jia-20 Datun Road, ChaoYang District, Beijing 100012, China}
\affiliation{State Key Laboratory of Radio Astronomy and Technology, Beijing 100101, China }
\email[notshow]{xujun@nao.cas.cn}  

\author[0000-0001-8241-1740]{Jian-Ning Fu}
\affiliation{Institute for Frontiers in Astronomy and Astrophysics, Beijing Normal University, Beijing 102206, People's Republic of China}
\affiliation{School of Physics and Astronomy, Beijing Normal University, Beijing 100875, People's Republic of China}
\email[notshow]{jnfu@bnu.edu.cn}

\begin{abstract}
The detailed evolution of triple star systems is complicated and poorly known. Using archived optical/infrared and $\gamma$-ray data, we identified the pulsar PSR~J0435+3233 as a $\gamma$-ray pulsar in a hierarchical triple system, with a helium white dwarf (WD) as a close inner binary companion and a Sun-like star as the distant tertiary. PSR~J0435+3233 and the WD companion are in a circular orbit with a period of $P_{\rm orb1} = 8.0$~days and an eccentricity of $e=0.00016$. The tertiary is a G-type subgiant with a mass of $0.98(12)\, M_\odot$ at a distance of $2.1(4)$\, kpc from the Earth. By simultaneously fitting the observed spin-period variations of the $\gamma$-ray emission (over 16.7 years) and radio emission (over 4.9 years) from PSR~J0435+3233, the changes in the inner orbital parameters, the Shapiro delay, Gaia astrometry, and the outer companion mass, we determined the outer elliptical orbit for the tertiary, with a period $P_{\rm orb2} = 73.5$~yrs and an eccentricity $e_2 = 0.598$. The outer orbit is either nearly perpendicular to the inner orbit with a mutual inclination of $\sim 85^\circ$, or moderately inclined by $\sim 54^\circ$. For the former geometry, the pulsar, the WD, and the tertiary star have masses of $1.16(5)\, M_\odot$, $0.273(8)\, M_\odot$, and $0.91(5)\, M_\odot$, respectively; for the latter geometry, the corresponding masses are $1.34(15)\, M_\odot$, $0.305(23)\, M_\odot$, and $1.13(7)\, M_\odot$. This is a unique triple system for detailed multi-band observations and for studying the evolutionary path and dynamic processes of a primordial triple star system. It will ultimately evolve into a system consisting of a neutron star and two white dwarfs. 
\end{abstract}


\section{Introduction}

Optical surveys suggested that a substantial fraction of intermediate- and high-mass stars reside in hierarchical triple systems \citep{Moe2017ApJS..230...15M}, with the triple fraction reaching approximately $20–30\%$ and even higher for the most massive stars. A hierarchical triple system can remain dynamically stable if the tertiary star orbits the inner binary on a sufficiently wide orbit \citep{Mardling2001MNRAS.321..398M}. Naturally, the massive stars in the triple system evolve and can produce black holes or neutron stars (NSs), while stars with intermediate or small masses produce white dwarfs (WDs). Therefore, some triple systems could have zero, one, two, or even three compact components, depending on the system's evolutionary stage. 

\begin{figure*}[tb]
    \centering
    \includegraphics[width=0.75\linewidth]{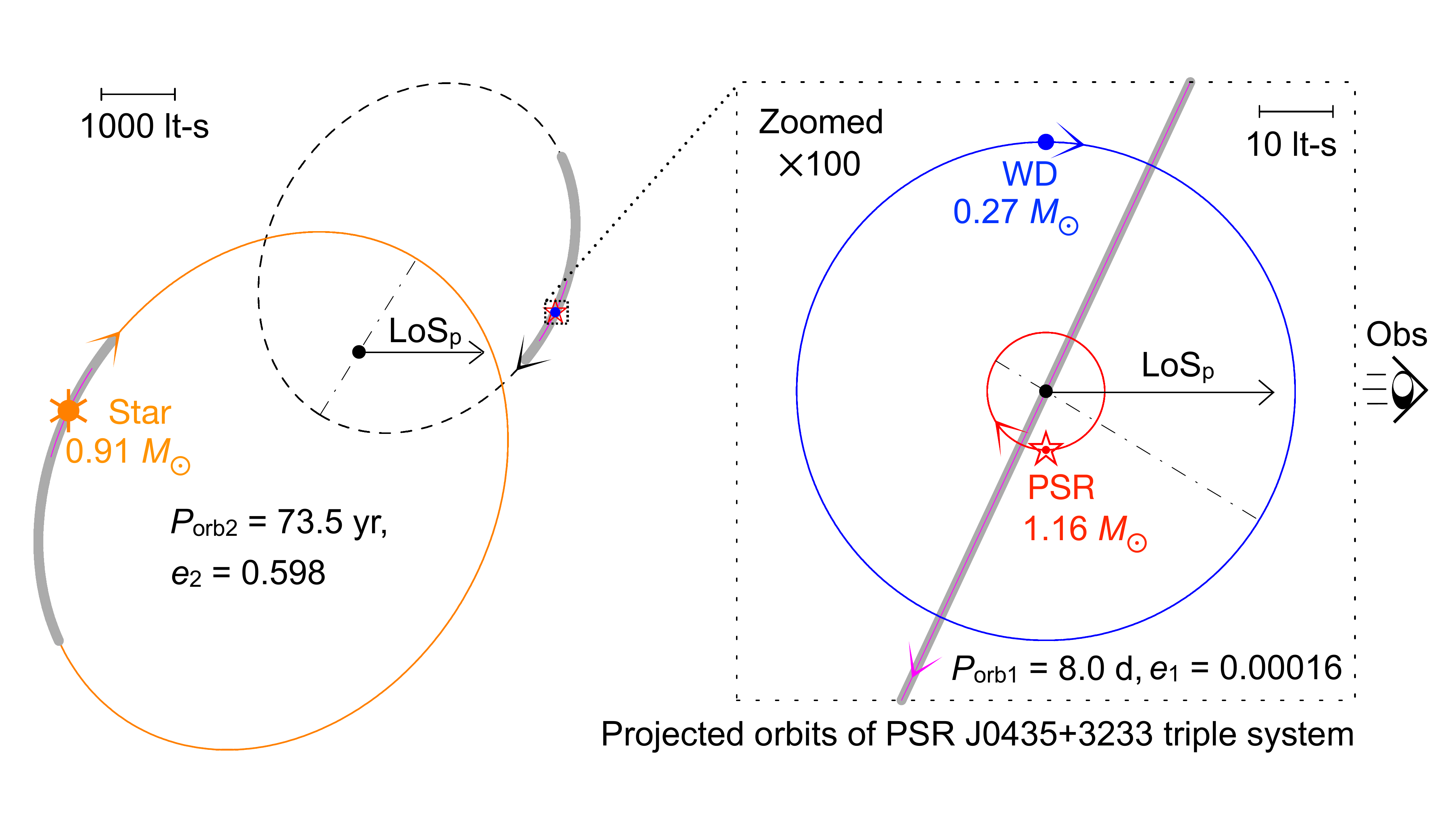}
    \caption{Projected orbits of PSR~J0435+3233 and its companions along the line of sight (based on the final timing solution A). The two black dots indicate the barycenters of the inner and outer orbits. The pulsar and the inner white-dwarf companion revolve clockwise around the inner barycenter with a period of 8 days, while the inner binary as a whole revolves clockwise around the triple-system barycenter with a period of about 73.5~yr. The line-of-sight (LoS$_{\rm p}$) is indicated by black arrows starting from the system's barycenters. The orbits shown here are the apparent orbits projected along the line of sight (the true orbital separation multiplied by $\sin i$). The red star marks the pulsar position at the epoch $T_{\rm asc1}=\text{MJD}~59997.48$ (as listed in Table~\ref{tab:FAST-timing-solution}), the blue dot marks the position of the inner white-dwarf companion, and the orange dot marks the position of the outer star companion. The two gray dotted-dashed lines connect the orbits' periastrons. The magenta line indicates the motion of the inner binary barycenter during the FAST data span, while the thick gray line indicates its motion during the long Fermi data span.}
    \label{Fig:geo}
\end{figure*}

Millisecond pulsars in gravitationally bound three-body systems are extremely rare. They are valuable samples for stellar evolution studies and also probes for tests of gravitational theories. PSR~B1620--26 \citep{Thorsett+1999ApJ...523..763T} is in a triple system, with the outer companion being an ultra-light object. PSR~J0337+1715 has two WD companions in two near-circular orbits 
\citep{Ransom+2014Natur.505..520R}. The relatively high masses of the two WD companions of PSR~J0337+1715 make gravitational three-body effects significant, and unique gravity tests have been performed based on the timing of this pulsar \citep{Shao+2016PhRvD..93h4023S, Iorio+2024Univ...10..206I}. 
The evolution of triple systems with pulsars has also drawn much attention. Many formation scenarios for PSR~J0337+1715 have also been proposed \citep{Tauris+2014ApJ...781L..13T, Voisin+2020AA...638A..24V, Sabach+2015MNRAS.450.1716S}. Further evolution of the PSR~J0337+1715 triple system can produce isolated millisecond pulsars or millisecond pulsars in eccentric orbits \citep{Portegies+2011ApJ...734...55P}.

The millisecond radio pulsar PSR J0435+3233, with a period of $P=3.20$~ms, was discovered by \citet{Wu+2026NatAs.tmp...71W} using the Five-hundred-meter Aperture Spherical radio Telescope  \citep[FAST;][]{Nan+2006ScChG..49..129N} during the Commensal Radio Astronomy FAST Survey \citep[CRAFTS;][]{Li+2018IMMag..19..112L}. After long-term efforts on pulsar timing, \citet{Wu+2026NatAs.tmp...71W} found that it has an extremely large period derivative compared to other known millisecond pulsars, and has been identified as a binary pulsar moving in an 8-day orbit, with a likely helium white dwarf (He WD) companion of mass $\sim0.3~M_\odot$. They also found that the orbital parameters of PSR~J0435+3233 vary rapidly.

We detected the $\gamma$-ray emission from PSR~J0435+3233 using the Fermi data \citep{Abdollahi+2022ApJS..260...53A} and identified a Sun-like G-type subgiant as its second companion, confirming that PSR~J0435+3233 is in a triple system comprising a $\gamma$-ray millisecond pulsar, a WD, and a Sun-like star (see Fig.~\ref{Fig:geo}). By accounting for the observed spin-period and orbital parameter variations of PSR~J0435+3233 induced by the gravitational perturbation from this tertiary, we determined the outer elliptical orbit, masses, and geometries. While preparing the submission of this paper, we noticed that \citet{Freire2026} worked independently on this object and obtained similar results, confirming each other well despite very different details. In the following, we present our data analyses and results, from the FAST radio data to the optical identification and long-term $\gamma$-ray data, and demonstrate how these data together constrain the properties of the PSR~J0435+3233 triple system.

\begin{table*}[bt]
	\centering
	\caption{FAST timing model parameters and derived quantities for PSR~J0435+3233. This FAST timing solutions are derived with \textsc{tempo2} \citep{Hobbs+2006MNRAS.369..655H} using the ELL1H binary model \citep{Lange+2001MNRAS.326..274L, Freire+2010MNRAS.409..199F} under the Barycentric Coordinate Time (TCB) system. Uncertainties quoted in parentheses correspond to the 1$\sigma$ errors for the last digit at the 68.3\% confidence level. }
    \label{tab:FAST-timing-solution}
\renewcommand\arraystretch{0.85}
\footnotesize
	\begin{tabular}{lr} 
\hline
		Parameter      & Value\\
\hline\\[-3mm]
 	   \multicolumn{2}{c}{General information}  \\     \hline                          
MJD range           &         59101 to 60877\\
Number of TOAs      &                 192 \\
EFAC value &  1.03 \\
Solar System ephemeris   &             DE440\\
Reference epoch (MJD)  &         59997.5\\
Binary model used   &      ELL1H\\
\hline\\[-3mm]
 	   \multicolumn{2}{c}{Model parameters}    \\  \hline 
Right ascension, $\alpha$ (J2000)   &      $04^{\rm h}35^{\rm m}33^{\rm s}.760877(3)$\\
Declination, $\delta$ (J2000)       &      $+32^\circ33'07''.93339(18)$\\
Proper motion in RA (mas\,yr$^{-1}$) & 1.31(4) \\
Proper motion in DEC (mas\,yr$^{-1}$) & -3.2(2) \\
Spin frequency, $\nu$ (Hz) &      312.710325717587(2)\\
1st spin frequency derivative, $\dot{\nu}$ (Hz $\rm s^{-1}$) & -4.7677041(2)$\times10^{-12}$\\
2nd spin frequency derivative, $\ddot{\nu}$ (Hz $\rm s^{-1}$) & -1.468823(2)$\times10^{-20}$\\
3rd spin frequency derivative, $\dddot{\nu}$ (Hz $\rm s^{-1}$) & -7.31287(10)$\times10^{-29}$\\
4th spin frequency derivative, $\nu^{(4)}$ (Hz $\rm s^{-1}$) & -4.5601(19)$\times10^{-37}$\\
5th spin frequency derivative, $\nu^{(5)}$ (Hz $\rm s^{-1}$) & -2.892(4)$\times10^{-48}$\\
6th spin frequency derivative, $\nu^{(6)}$ (Hz $\rm s^{-1}$) & -9.0(9)$\times10^{-54}$\\
Dispersion measure, DM (pc\,cm$^{-3}$) & 37.32427(7) \\
Orbital frequency, $f_{\rm orb1}$ (Hz) & 1.44709060570(7)$\times$10$^{-6}$ \\
1st derivative of $f_{\rm orb1}$, $\dot{f}_{\rm orb1}$ (s$^{-2}$) & -2.2251(5)$\times10^{-20}$ \\
2nd derivative of $f_{\rm orb1}$, $\ddot{f}_{\rm orb}$ (s$^{-3}$) & -7.054(12)$\times10^{-29}$  \\
3rd derivative of $f_{\rm orb1}$, $\dddot{f}_{\rm orb1}$ (s$^{-4}$) & -3.79(13)$\times10^{-37}$ \\
Projected semi-major axis, $x_1$ (lt-s) & 7.97814982(10) \\
1st derivative of $x_1$, $\dot{x}_1$ (lt-s\,s$^{-1}$) & 6.043(17)$\times10^{-13}$ \\
2nd derivative of $x_1$, $\ddot{x}_1$ (lt-s\,s$^{-2}$) & 1.87(8)$\times10^{-22}$\\
Time of ascending node passage, $T_{\rm asc1}$ (MJD) & 59997.477140343(15) \\
First Laplace parameter, $\epsilon_1=e_1\sin\omega_1$   &  0.000139437(15) \\
Second Laplace parameter, $\epsilon_2=e_1\cos\omega_1$ &  -0.000084127(14) \\
3rd orthometric amplitude, $h_3$ ({\textmu}s) & 0.44(4)\\
4th orthometric amplitude, $h_4$ ({\textmu}s) & 0.34(6)\\
\hline\\[-3mm]
 	   \multicolumn{2}{c}{Derived parameters}    \\ 
\hline
Spin period, $P$ (s) & 0.00319784771323193(3) \\
Orbital period, $P_{\rm orb1}$ (day) & 7.9981682063(4) \\ 
Orbital eccentricity, $e_1$ & 0.000162850(15) \\
Longitude of periastron, $\omega_1$ (deg) &
121.104(5) \\
Mass function ($M_\odot$) & 0.0085233088(3) \\
\hline 
\hline
    \end{tabular}
\end{table*}

\section{Identification of the PSR J0435+3233 triple system}

We identified the PSR J0435+3233 triple system from the FAST radio data, archived optical/infrared images, and Fermi data.

\subsection{Reprocessing the FAST archive data and timing PSR J0435+3233}
  
PSR J0435+3233 has been observed by FAST for 192 sessions in the last 5 years \citep{Wu+2026NatAs.tmp...71W} using the 19-beam L-band receiver \citep{Jiang+2020RAA....20...64J}. All FAST observations of PSR J0435+3233 cover a frequency range of 1.0 to 1.5~GHz, and the data of all sessions were recorded in four polarization channels: $XX$, $YY$, $\text{Re}[XY]$, and $\text{Im}[XY]$. At the beginning of most observations, the calibration noise signals with an amplitude of 12.5~K and a period of 0.100663\,s were injected into the receiver feeds for 40\,s, thereby enabling polarization calibration.

\begin{figure}[tb]
    \centering
    \includegraphics[width=0.93\linewidth]{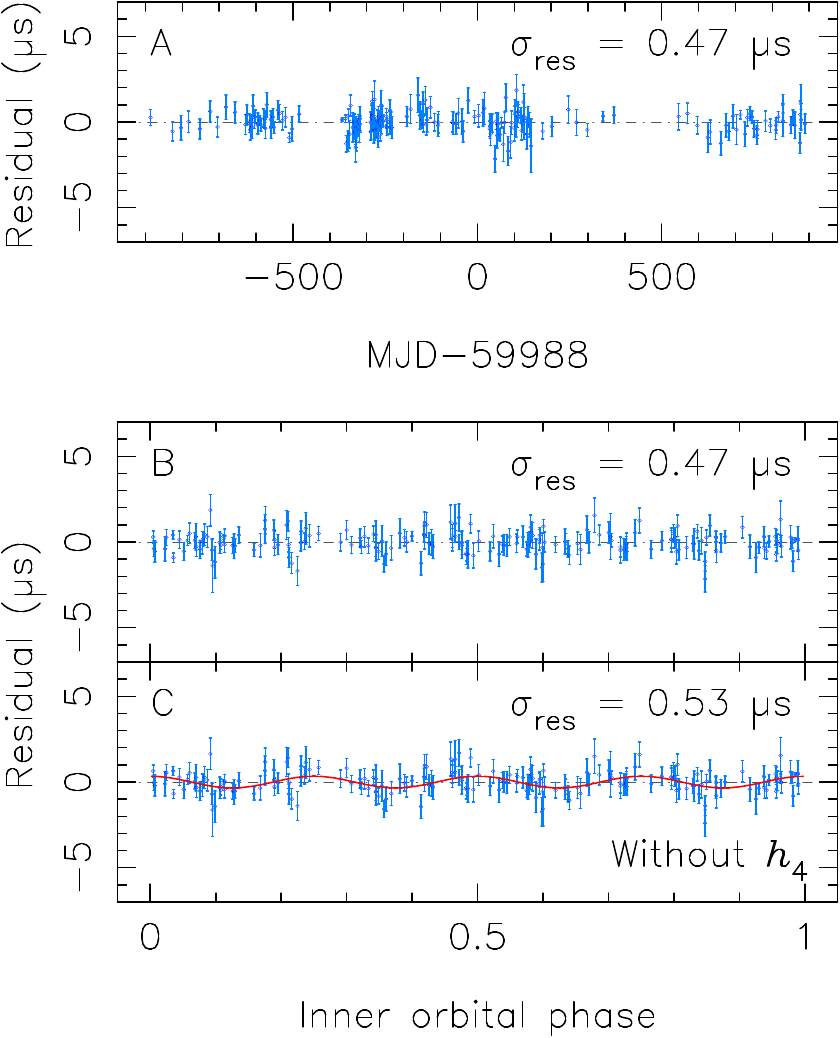}
    \caption{FAST timing residuals of PSR J0435+3233 based on the ELL1H model (see Table~\ref{tab:FAST-timing-solution}). Panel ({\bf A}) shows the timing residuals across the observation dates (in MJD). Panels ({\bf B}) and {\bf C}) show the residuals as a function of inner orbital phase (measured from the ascending node passage) with and without considering the Shapiro delay parameter $h_4$ in the timing model. 
    The corresponding 4th harmonic terms are also indicated by the red line. The expected peak phase of the Shapiro-delay term is 0.25 when all harmonics are included. 
    }
    \label{fig:res}
\end{figure}

We reprocessed the released FAST archive data of 192 sessions in the MJD range from 59101 to 60877 for timing PSR J0435+3233. We folded the raw data using {\sc dspsr} \citep{dspsr2011}, producing pulse polarization profiles with 128 phase bins across the pulse longitude. Polarization calibration was then applied via the \texttt{pac} routine within the \textsc{psrchive} package \citep{Hotan+2004PASA...21..302H}. For each session, we obtained the polarization profiles for 4 frequency subbands. The pulsar shows strong scintillation. We built the standard profile using the \texttt{paas} command \citep{Hotan+2004PASA...21..302H} from the observation with the highest flux density. Then we extracted pulse times of arrival (TOAs) with the \texttt{pat} command \citep{Hotan+2004PASA...21..302H}.
We refined the DM value by comparing TOAs in four subbands, then extracted TOAs from integrated profiles of 192 sessions.
Based on the derived TOAs, we performed pulsar timing analysis using the \textsc{tempo2} software package \citep{Hobbs+2006MNRAS.369..655H}. We shifted the spin-frequency reference epoch to MJD 59997.5, close to the ascending-node passage time $T_{\rm asc}$ (adopted as the orbital-frequency reference epoch). 
The TOA uncertainties have been multiplied by a factor EFAC to set the reduced $\chi^2$ value to 1. 
Table~\ref{tab:FAST-timing-solution} lists the final timing solutions and derived parameters, and Fig.~\ref{fig:res} plots the corresponding timing residuals. 
Our timing results are consistent with those of \citet{Wu+2026NatAs.tmp...71W}, but show notable improvements in residuals and parameter accuracy.  

The pulsar's close companion in the inner orbit is a WD. With the total mass of the inner binary, $M_{\rm in}=M_{\rm PSR}+M_{\rm WD}$, the mass function of the two compact objects in the inner orbit is
\begin{equation}
\begin{aligned}
    f (M_{\rm in},M_{\rm WD},i_1)
    =\frac{(M_{\rm WD} \sin i_1)^3}{M_{\rm in}^2}=\frac{4\pi^2}{G}\frac{x_1^3}{P_{\rm orb1}^2} \\ =0.0085233093(2)\,M_\odot,
    \label{eq:mf_in}
\end{aligned}
\end{equation} 
where $G$ is the gravitational constant. 

In the ELL1H binary model \citep{Freire+2010MNRAS.409..199F}, the Shapiro delay term is also expressed by a number of harmonics of the orbital period. In the triple system, however, we cannot measure the first two harmonics because they are coupled to the orbital parameters, while the outer companion's perturbation can introduce a systematic offset in the third harmonic $h_3$ (see below). 
Only the fourth harmonic $h_4$ is the smallest harmonic and free of perturbed bias, so it can be used for mass constraints.
Based on the ELL1H binary model \citep{Freire+2010MNRAS.409..199F} with some modifications to include the high-order orbital parameter derivatives, the Shapiro delay has been detected for the inner orbit in our timing analysis, yielding $h_3 = 0.44(4)$\,\textmu s and $h_4=0.34(6)$. In general relativity,
\begin{equation}
\begin{aligned}
    h_3&=\frac{G M_{\rm WD}}{c^3}\varsigma^3,\\
    h_4&=\frac{G M_{\rm WD}}{c^3}\varsigma^4,\\
    \varsigma&=(\frac{\sin i_1}{1+|\cos i_1|})^{1/2},
\end{aligned}
    \label{eq:Shapiro}
\end{equation}
where $c$ is the speed of light. Hereafter, we take $c=1$ and omit it in the following formula. The orbital inclination $i_1$ remains ambiguous because the standard Shapiro delay cannot distinguish between $i_1$ and $180^\circ-i_1$. 
From these values, we obtained $M_{\rm WD}= \varsigma^{-4}\times0.069(12)\,M_\odot$. Combining $h_4$ and the mass function, and assuming a typical pulsar mass of $M_{\rm PSR}=1.4\, M_\odot$, we find an inclination angle for the orbital plane of $i_1\sim69^\circ$ (or $111^\circ$) and a companion mass of $M_{\rm WD}\sim0.31\, M_\odot$. Based on previous measurements \citep{Alsing+2018MNRAS.478.1377A, Muller+2025PhRvL.134g1403M}, NS masses are most likely to lie between 1.1 and 2.2\,$M_\odot$. For the minimum pulsar mass of $M_{\rm PSR}=1.1\,M_\odot$, the inclination angle should be $i_1=74^\circ$ (or $106^\circ$) and the companion mass be $M_{\rm WD}=0.26\,M_\odot$; for the maximum pulsar mass of $M_{\rm PSR}=2.2\,M_\odot$, the inclination angle should be $i_1=66^\circ$ (or $114^\circ$) and the companion mass is $M_{\rm WD}=0.43\,M_\odot$. Therefore, the total mass of the inner binary should be between 1.4 and 2.6\,$M_\odot$, and the inclination angle of the inner orbit should be $i_1 \sim 70^\circ$ or 110$^\circ$.

Millisecond pulsars with He white-dwarf companions that descend from low-mass X-ray binaries follow a $P_{\rm orb}$--$M_{\rm WD}$ relation \citep{Tauris+1999A&A...350..928T}
\begin{equation}
    \frac{M_{\rm WD}}{M_\odot} = \left(\frac{P_{\rm orb}}{1.1\times10^5~{\rm days}}\right)^{1/4.75} + 0.115.
\end{equation}
For the corresponding orbital period in Table~\ref{tab:FAST-timing-solution}, the derived He WD mass is 0.25~${M_\odot}$, not much different from the above-derived mass. 

Furthermore, we successfully detected the second derivative of the projected semi-major axis, $\ddot{x}$.
After shifting the reference epoch of the orbital frequency and spin frequency to MJD 59997.5 (see Table~\ref{tab:FAST-timing-solution}), we found that the values of $\dot{\nu}/\nu=-1.5246\times10^{-14}~{\rm s}^{-1}$ and $\dot{f}_{\rm orb1}/{f}_{\rm orb1}=-1.5382(3)\times10^{-14}~{\rm s}^{-1}$ are consistent up to 99\%, indicating that the variations are caused by the tertiary in this system. 

\begin{table*}[tb]
    \centering
    \caption{Optical/infrared parameters for the tertiary. The geometry information, apparent magnitudes in different bands, and derived parameters for Gaia DR3 171833395176138240, which is exactly at the timing position of PSR J0435+3233. Numbers in parentheses represent 1$\sigma$ errors in the final decimal.} 
    \footnotesize 
    \renewcommand\arraystretch{0.85}
	\setlength{\tabcolsep}{2mm}
    \begin{tabular}{llc}
    \hline
    Parameter & Value & Reference \\
    \hline
    \multicolumn{3}{c}{Geometric parameters} \\ 
    \hline 
    Right ascension, $\alpha$ (J2000) &  $04^{\rm h}35^{\rm m}33^{\rm s}.761275(4)$ & \citep{Gaia+2023AA...674A...1G} \\
    Declination, $\delta$ (J2000)    & $+32^\circ33'07''.95514(4)$ & \citep{Gaia+2023AA...674A...1G} \\
    Reference epoch & J2016.0 & \citep{Gaia+2023AA...674A...1G} \\
    Proper motion in RA (mas\,yr$^{-1}$)  & 0.694(84) & \citep{Gaia+2023AA...674A...1G} \\
    Proper motion in DEC (mas\,yr$^{-1}$)  & -3.530(63) & \citep{Gaia+2023AA...674A...1G} \\
    Parallax (mas) & 0.466(78) & \citep{Gaia+2023AA...674A...1G} \\
    \hline
        \multicolumn{3}{c}{Apparent magnitudes} \\ \hline
    PS1 $g$ (mag) & 17.4601(42) & \citep{Chambers+2016arXiv161205560C}  \\
    Gaia DR3 $BP$ (mag) & 17.2978(63) & \citep{Gaia+2023AA...674A...1G} \\
    PS1 $r$ (mag) & 16.6684(24) & \citep{Chambers+2016arXiv161205560C}  \\
    Gaia DR3 $G$ (mag) & 16.6539(28) & \citep{Gaia+2023AA...674A...1G} \\
    PS1 $i$ (mag) & 16.2699(20) & \citep{Chambers+2016arXiv161205560C} \\
    Gaia DR3 $RP$ (mag) & 15.8756(51) & \citep{Gaia+2023AA...674A...1G} \\
    TESS & 15.9447(98) & \citep{Ricker+2015JATIS...1a4003R}\\
    PS1 $z$ (mag) & 16.0625(33) & \citep{Chambers+2016arXiv161205560C}  \\  
    PS1 $y$ (mag) & 15.9178(56) & \citep{Chambers+2016arXiv161205560C}  \\
    2MASS $J$ (mag) & 14.847(39) & \citep{Skrutskie+2006AJ....131.1163S} \\
    2MASS $H$ (mag) & 14.515(46) & \citep{Skrutskie+2006AJ....131.1163S} \\
    2MASS $K_{\rm s}$ (mag) & 14.302(65) & \citep{Skrutskie+2006AJ....131.1163S} \\
    WISE $W1$ (mag) & 14.256(59) & \citep{Skrutskie+2006AJ....131.1163S} \\   WISE $W2$ (mag) & 14.277(70) & \citep{Skrutskie+2006AJ....131.1163S} \\ 
    \hline
    \multicolumn{3}{c}{SHBoost catalog parameters} \\
    \hline
    Surface gravity, $\log_{10}(g)$ & $4.18(20)$ & \citep{Khalatyan+2024AA...691A..98K} \\
    Stellar mass, $M_{\rm star}$ ($M_\odot$) & 0.98(12) & \citep{Khalatyan+2024AA...691A..98K} \\ 
    \hline
    \end{tabular}
    \label{tab:optical}
\end{table*}

\begin{figure}
    \centering
    \begin{tikzpicture}
        \node[inner sep=0] (img)
            {\includegraphics[width=0.43\linewidth]{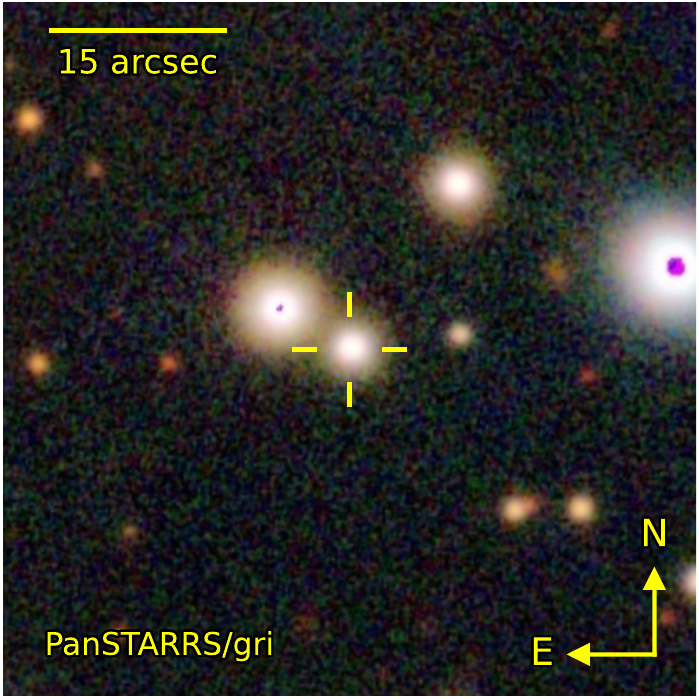}};
        \node[anchor=south east, font=\bfseries\footnotesize,
              xshift=0.2em, yshift=-0.5em] at (img.north west) {A};
    \end{tikzpicture}
    \begin{tikzpicture}
        \node[inner sep=0] (img)
            {\includegraphics[width=0.43\linewidth]{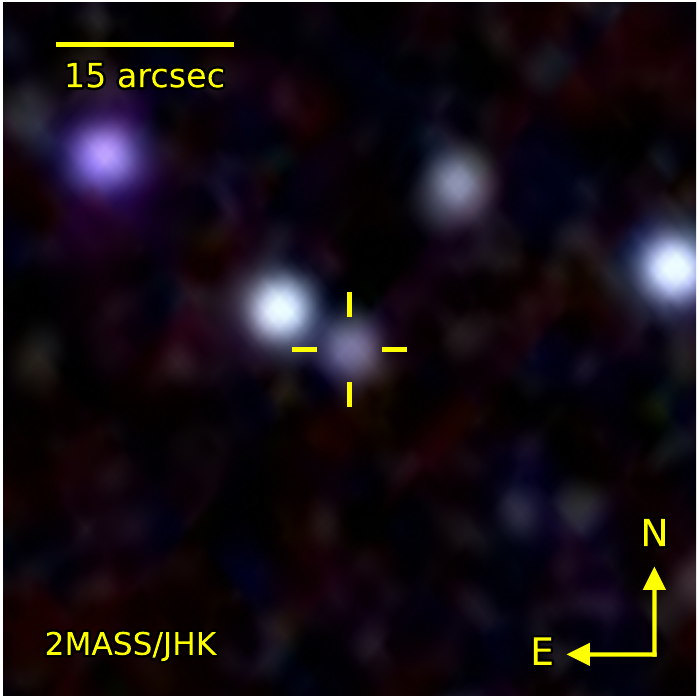}};
        \node[anchor=south east, font=\bfseries\footnotesize,
              xshift=0.2em, yshift=-0.5em] at (img.north west) {B};
    \end{tikzpicture}
    \caption{Optical and infrared counterpart of the outer companion of PSR~J0435+3233. In the false-color image of PanSTARRS \citep{Chambers+2016arXiv161205560C} and 2MASS \citep{Skrutskie+2006AJ....131.1163S}, a source is found at the location of PSR~J0435+3233, which is the counterpart of its outer companion.}
    \label{fig:image}
\end{figure}

\subsection{Identifying the optical/infrared counterpart as the tertiary}

At the almost exact position of PSR J0435+3233 (see Table~\ref{tab:FAST-timing-solution}), we found an optical and infrared counterpart (see Fig.~\ref{fig:image}), Gaia DR3 171833395176138240, from the archived images or catalogues of the infrared and optical bands by the Pan-STARRS \citep{Chambers+2016arXiv161205560C}, Gaia \citep{Gaia+2016AA...595A...1G}, the TESS \citep{Ricker+2015JATIS...1a4003R}, 2MASS \citep{Skrutskie+2006AJ....131.1163S}, and WISE \citep{Wright+2010AJ....140.1868W}. 
According to the Gaia DR3 \citep{Gaia+2023AA...674A...1G} astrometry, at the reference epoch at J2016.0 (MJD 57388.5), the star is located at the right ascension $04^{\rm h}35^{\rm m}33^{\rm s}.761275(4)$ and the declination $+32^\circ33'07''.95514(4)$, with a proper motion of $0.694(84)$~mas~yr$^{-1}$ in the right ascension and $-3.530(63)$~mas~yr$^{-1}$ in the declination, consistent with the proper motion of the pulsar determined from the FAST timing observations (for the given the uncertainties). 

Using measurements from these optical/infrared surveys, we can further determine the star's properties. 
The parallax of the star measured Gaia DR3 \citep{Gaia+2023AA...674A...1G} is $0.466(78)$~mas, corresponding to a distance of $D = 2.1(4)$ kpc. At this distance, the star has a bolometric luminosity slightly greater than the Sun, with an effective temperature similar to the Sun, suggesting a subgiant classification. 
In the catalog of stellar properties derived from Gaia DR3 XP spectra \cite[SHBoost,][]{Khalatyan+2024AA...691A..98K}, its mass is estimated to be $M_{\rm star} = 0.98(12)\,M_\odot$, and a surface gravity\footnote{Note that the value of $\log_{10}(g)=4.68$ in the original Gaia catalog DR3 \citep{Gaia+2023AA...674A...1G} probably includes the contamination from a nearby luminous source about $7''$ away.} to be $\log_{10}(g)=4.18(20)$, confirming that it is a G1V star, a sub-giant with a temperature of 5854~K and luminosity of 2.1~$L\odot$. Table~\ref{tab:optical} lists the properties of this tertiary star.

\begin{figure}[t]
    \centering
    \begin{tikzpicture}
        \node[inner sep=0] (img)
            {\includegraphics[width=0.8\linewidth]{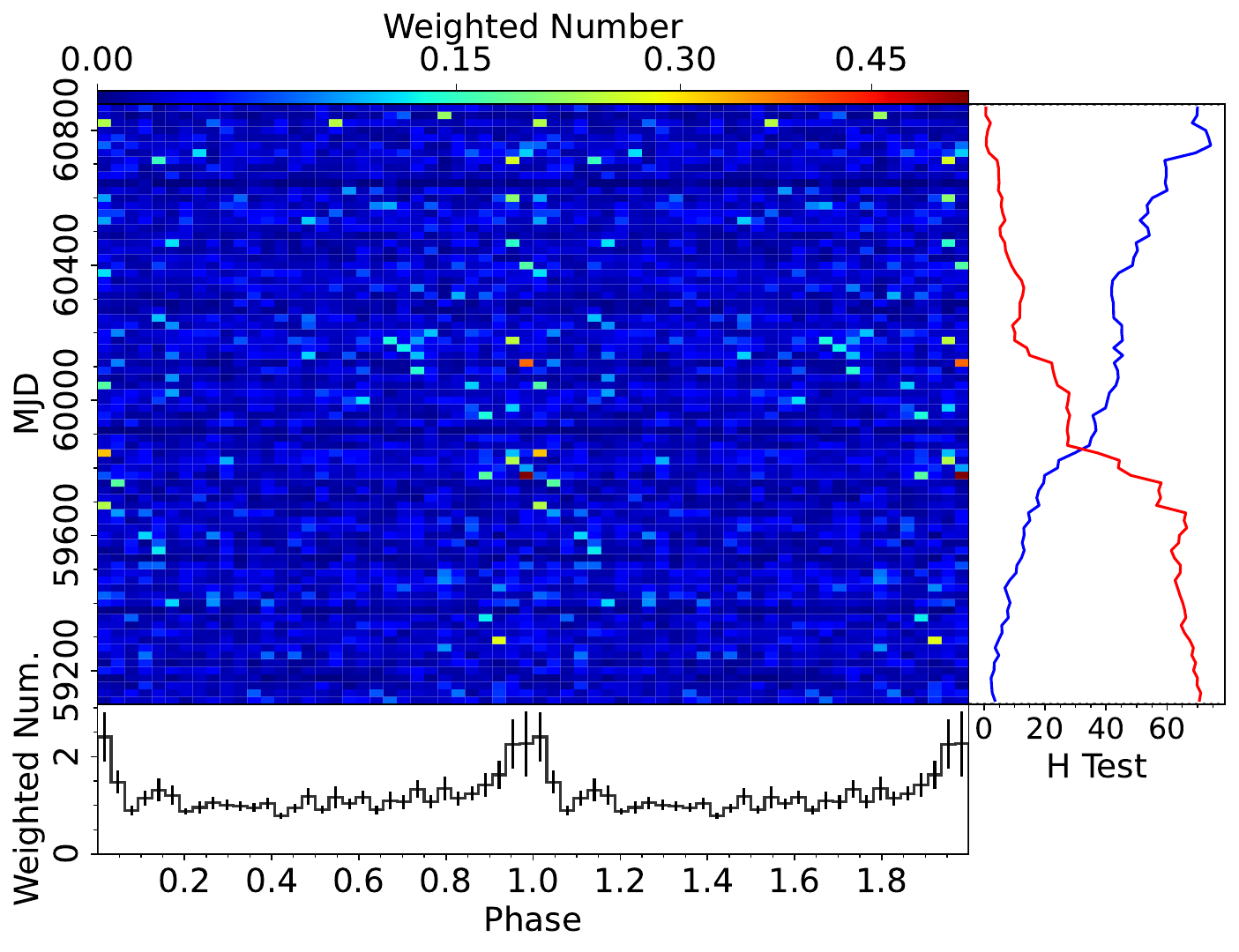}
            };
        \node[anchor=south east, font=\bfseries\footnotesize,
              xshift=1.2em, yshift=-1.4em] at (img.north west) {A};
    \end{tikzpicture}
    \begin{tikzpicture}
        \node[inner sep=0] (img)
            {\includegraphics[width=0.8\linewidth]{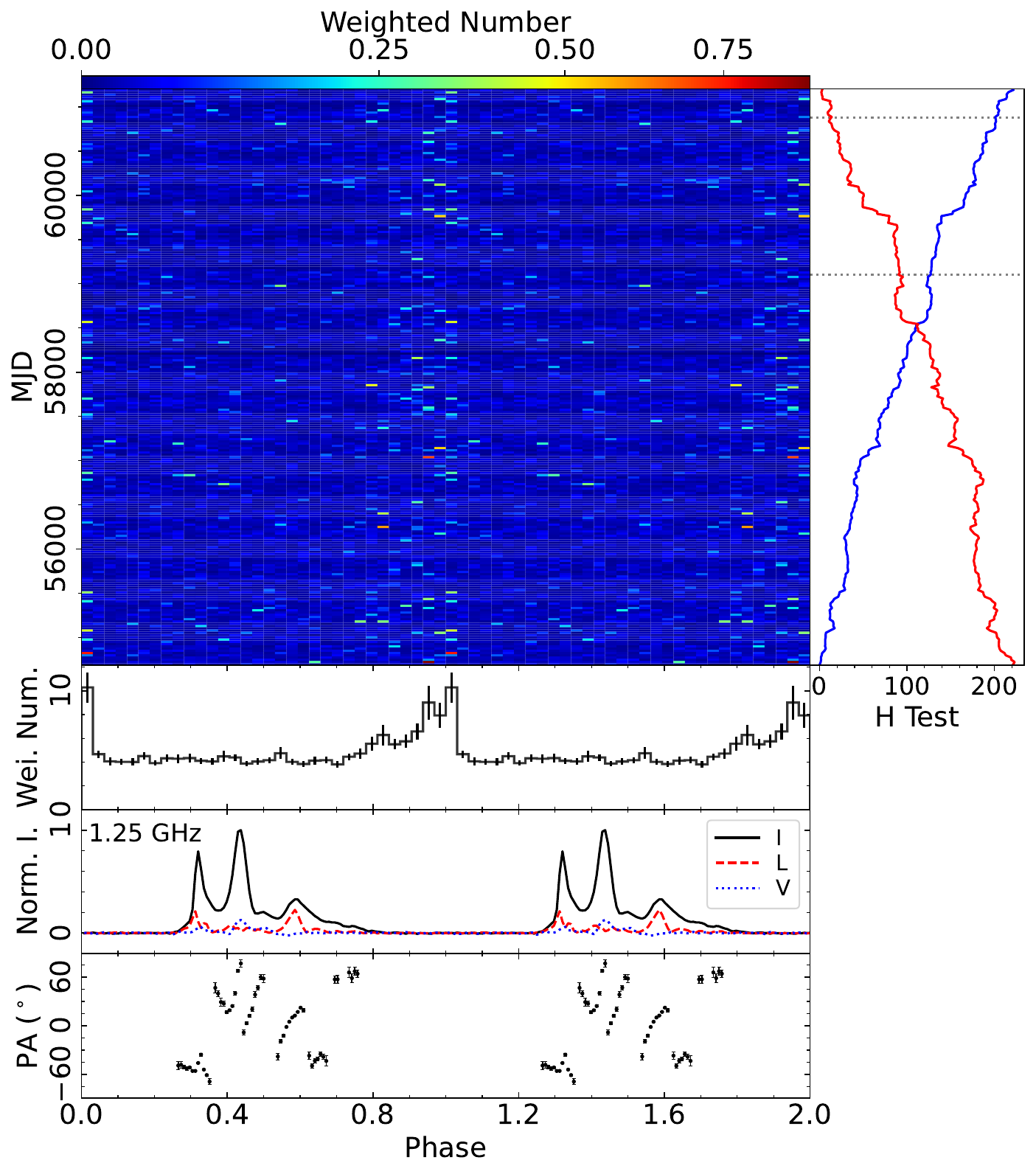
            }};
        \node[anchor=south east, font=\bfseries\footnotesize,
              xshift=1.2em, yshift=-1.4em] at (img.north west) {B};
    \end{tikzpicture}
    \caption{\small Phase histograms of $\gamma$-ray photons for PSR J0435+3233 and aligned with radio polarization profiles. 
    Panel (A) is the result with the H-test value of $H \simeq$  69 for the Fermi data in the FAST observation duration (MJD: 59101 -- 60877) and folded using the timing solution A with $i_{\rm x}\sim85^\circ$ (see Table~\ref{tab:timing_solution_final}). Note that folding Fermi data from MJD 59101 to 60783 yields $H \simeq73$. Panel (B) shows the result from all available Fermi data (MJD: 54684 -- 61205 $\simeq 18$~years) using the final improved timing solution A, with the $H$ test factor reaching  221.  
    The folding results of the timing solution B with $i_{\rm x}\sim54^\circ$ give the similar H-test values.
    The error bars of the Fermi profile show the 1$\sigma$ uncertainties, which are the square roots of the sum of squared weights.
 The lower two panels show timing-aligned radio polarization profiles at 1.25 GHz, showing a half-period offset from the Fermi profile. The total intensity $I$ (solid line), linearly polarized intensity $L$ (dashed line), and circularly polarized emission $V$ (dotted line), with positive values for the left-hand sense, were obtained from 20-min FAST observations on 2022 August 6, which has the highest signal-to-noise ratio. The bottom subpanel plots the linear polarization position angles (PAs), with uncertainties of $\pm 1\sigma$.  
    }
    \label{fig:gammaFoldplot1}
\end{figure}

We calculated the Gaia tertiary star coordinates in Table~\ref{tab:optical} to the pulsar's reference epoch (MJD 59997.5) using its measured proper motions, then compared them with the pulsar's timing-determined position, yielding an offset of 10~mas in right ascension and $-3$~mas in declination. Based on the distance to the tertiary star inferred from the parallax, $2.1(4)$~kpc, and the offsets, the projected separation of the tertiary star and the pulsar in the sky plane can be estimated to be $S_{\perp} = 1.1(2)\times10^4$~lt-s. Because their relative proper motions are $-0.82(9)$~mas\,yr$^{-1}$ in the right ascension and $-0.1(2)$~mas\,yr$^{-1}$ in the declination, their projected separation on the sky plane is decreasing, implying that the pulsar and the tertiary star are approaching each other. Note that the orbital motion of the tertiary star may lead to a systematic error in parallax measurement. However, for a not-so-wide phase range of the orbit, the proper motions and parallax measured by Gaia still reasonably describe the tertiary star's motion. A better description that accounts for orbital motion will be desired when the observational time span is comparable to the outer orbital period. 

\subsection{Initial detection of the $\gamma$-ray emission of PSR J0435+3233}

We found that a $\gamma$-ray source, 4FGL J0435.5+3232, in the 4th Fermi LAT point source catalog \citep{Abdollahi+2022ApJS..260...53A} is only 0.8~arcminute away from PSR J0435+3233. To verify whether this $\gamma$-ray source is the pulsar's counterpart, we fold the Fermi data using the ephemeris obtained from the FAST observations (Table~\ref{tab:FAST-timing-solution}). 
We extracted LAT photons with energies between 100 MeV and 300 GeV collected during the FAST timing span from the sky area within a 2$^\circ$ radius of the pulsar coordinates. Based on the radio pulsar ephemeris, we used the \textsc{tempo2} Fermi plug-in to compute the phases of the photon arrival times \citep{Hobbs+2006MNRAS.369..655H, Edwards2006, Ray+2011ApJS..194...17R}. Following the simple weight method of \citet{Bruel2019} to calculate the photon weights, we folded the Fermi photon data in the FAST observation time span (MJD range 59101 to 60877) using the timing parameters in Table~\ref{tab:FAST-timing-solution}, and found that PSR J0435+3233 exhibits $\gamma$-ray pulsations, with an H-test value of 69, confirming the pulsar's $\gamma$-ray emission (see Fig.~\ref{fig:gammaFoldplot1}A). 
While we were preparing the manuscript, we noticed that \citet{zhang2026Fermi} obtained a similar result (with an $H$ value of 63.9).

We show below that better-determined triple-system parameters improve the pulsar timing solution and extend the Fermi data span, yielding a much improved $\gamma$-ray profile for this millisecond pulsar. 

\begin{figure*}[tb]
    \centering
    \includegraphics[width=0.42\linewidth]{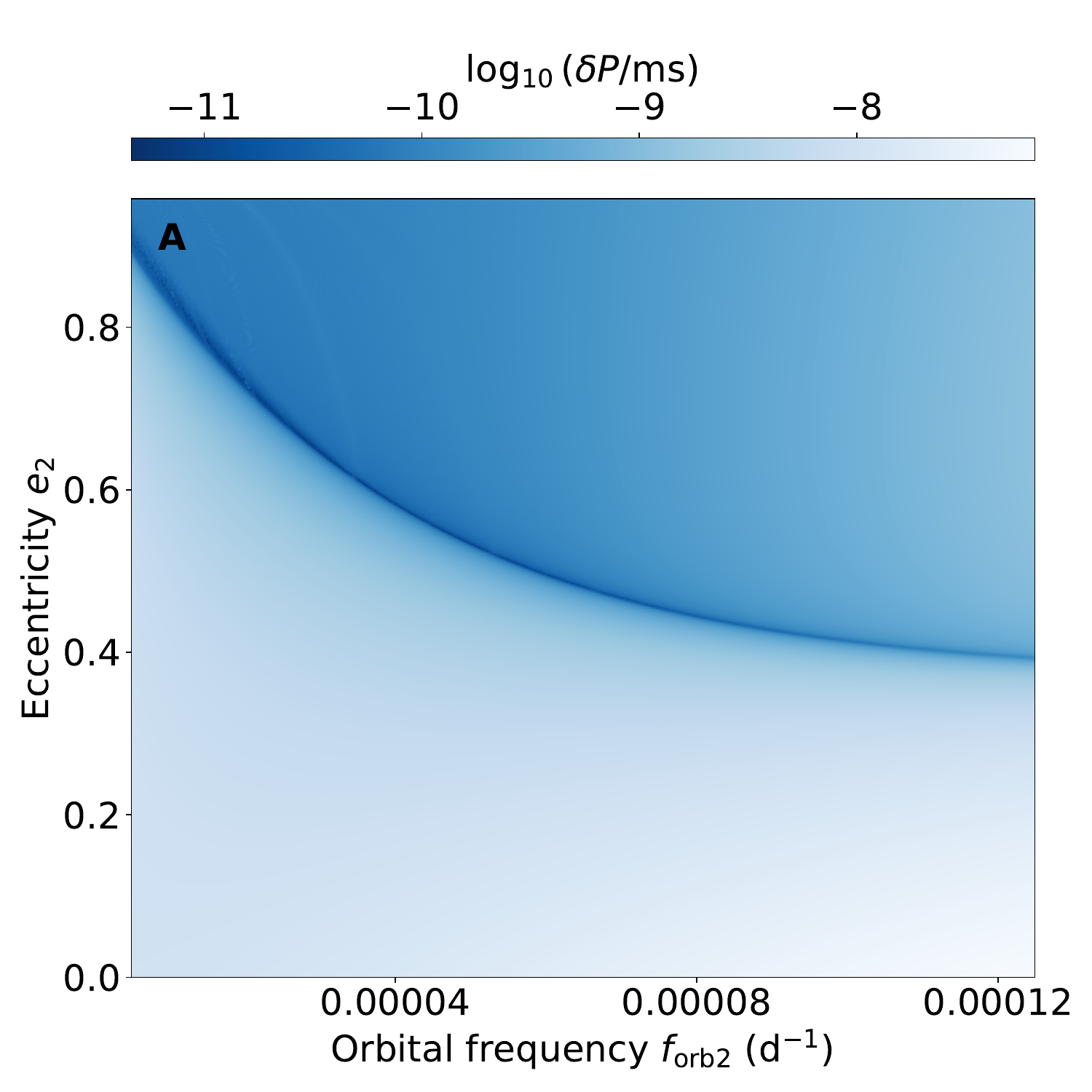}
    \includegraphics[width=0.40\linewidth]{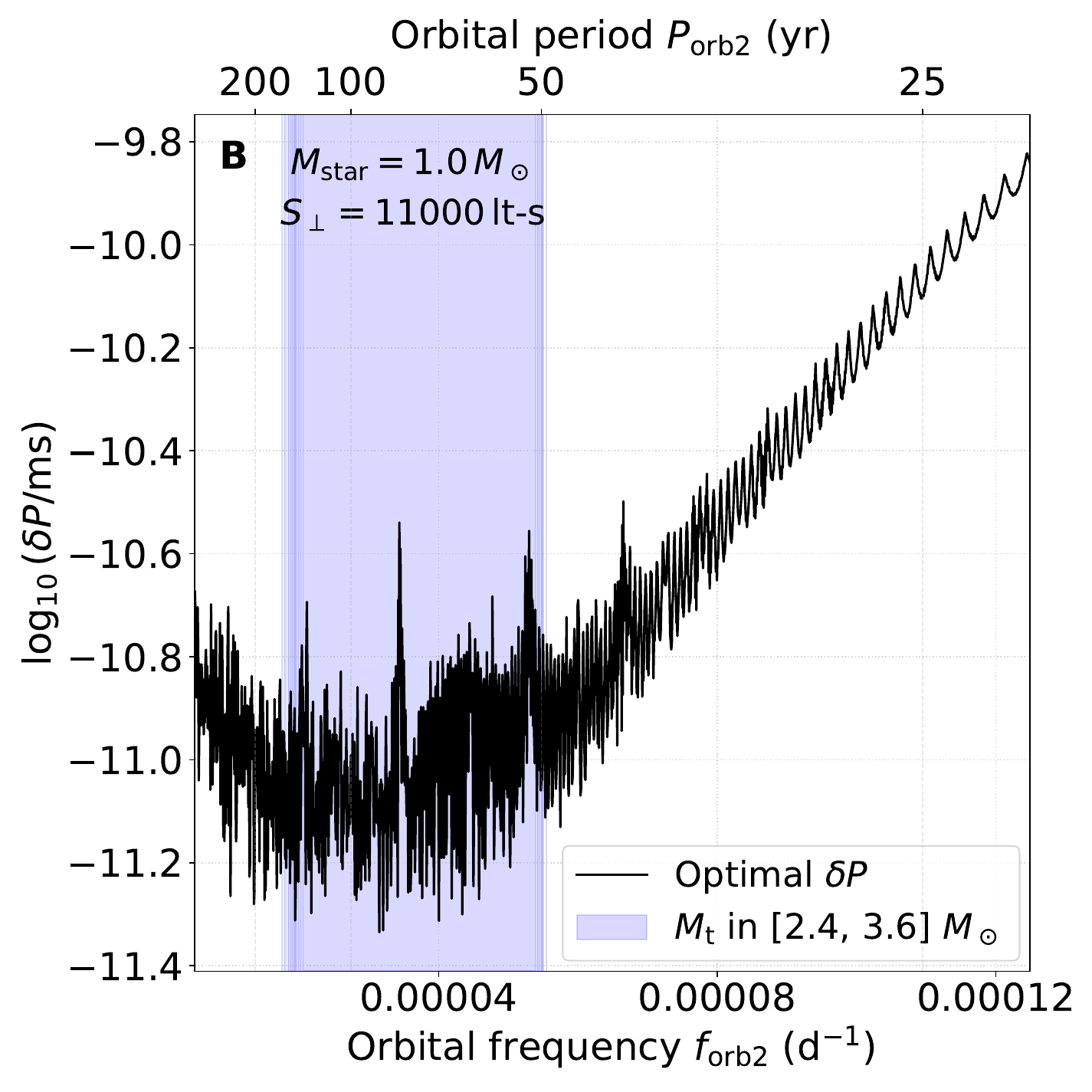}
    \caption{Preliminary constraints on the outer orbital parameters.
    Panel~(A): Residual distribution of the FAST-measured pulsar periods ($\delta P$ in ms) versus eccentricity $e_2$ and orbital frequency $f_{\rm orb2}$ (yr$^{-1}$), plotted on a logarithmic scale. The best fits follow the darker curve, which shows the coupling between $e_2$ and $f_{\rm orb2}$.
    Panel~(B): Best-fit $\delta P$ as a function of $f_{\rm orb2}$ for a given set of $M_{\rm star}=1~M_\odot$ and $S_{\perp}=11000$~lt-s constrained from the optical and infrared counterpart. The total mass $M_{\rm t}$ for the optimal orbital parameters should be in the range $2.4$--$3.6\, M_\odot$, with $f_{\rm orb2}$ highlighted by the shadowed region, corresponding to the orbital period of the outer orbit $P_{\rm orb2}$ between 50 and 150 yr and the eccentricity $e_2$ between 0.50 and 0.75. 
    }
    \label{Fig:fit}
\end{figure*}

\section{Constraints on the properties of the triple system}

Since we have identified the tertiary star, the total mass of the triple system is $M_{\rm t} \equiv M_{\rm PSR}+M_{\rm WD}+M_{\rm star}$. Using FAST, optical/infrared, and Gamma-ray data, we aim to determine the parameters of the inner and outer orbits. 

\subsection{From observed pulsar period variations by FAST}

The orbital parameters of the outer orbit can be constrained using the observed pulsar period variations caused by the orbital motion, which can be reconstructed from the pulsar timing results in Table~\ref{tab:FAST-timing-solution} via
\begin{equation}
    1/P(t)\equiv\nu(t)=\nu+(\dot{\nu}-\dot{\nu}_0)(t-t_0)+\frac{1}{2}\ddot\nu(t-t_0)^2+...,
\end{equation}
where $t_0$ denotes the reference epoch, $\dot{\nu}_0$ accounts for the intrinsic spin-down, and $t$ is the coordinate time of emission, which can be approximated by the infinite-frequency barycentre arrival time $t_{\rm b}$. We list the first five spin-frequency derivatives in Table~\ref{tab:FAST-timing-solution}, obtained by fitting the FAST observational data. The intrinsic spin-down $\dot{\nu}_0$ is largely coupled into the pulsar motions caused by orbital motion and cannot be directly determined. 

The pulsar spin period varies with the outer orbital phase and can be expressed in terms of the orbital parameters by 
\begin{equation}
\begin{aligned}
    P 
    &=P_0\left( 1+\frac{{\rm d}\Delta_{\rm R2}}{{\rm d}t} \right)\\&= P_0+P_0\frac{2\pi x_2f_{\rm orb2}}{\sqrt{1-e_2^2}}[(\cos \theta_2+e_2)\cos\omega_2\\
    &~~~~~~~~~~~~~~~~~~~~~~~~~~~~~~~~~~~-\sin\theta_2\sin\omega_2] ,
\end{aligned}
\end{equation}
where $P_0$ is the intrinsic spin period of the pulsar, $\Delta_{\rm R2}$ is the R\"{o}mer delay, $\theta_2$ is the true anomaly, $e_2$ is the eccentricity, and $\omega_2$ is the longitude of the periastron of the outer orbit. Here, $\theta_2$ is related to the outer orbital frequency $f_{\rm orb2}\equiv1/P_{\rm orb2}$ and the epoch of periastron passage of the outer orbit $T_{\rm 02}$ via
\begin{equation}
\begin{aligned}
    E_2-e_2\sin E_2 &=2\pi f_{\rm orb2}(t-T_{\rm 02}),\\
    \cos \theta_2 &= \frac{\cos E_2-e_2}{1-e_2\cos E_2},\\
    \sin \theta_2 &= \frac{\sqrt{1-e_2^2} \sin E_2} {1 - e_2 \cos E_2}.
\end{aligned}
\end{equation}
Here $E_2$ is the eccentric anomaly of the outer orbit. Since the period $P$ depends linearly on $(\cos \theta_2+e_2)\cos\omega_2-\sin\theta_2\sin\omega_2$, one can use linear regression to fit the model and quantify fit quality by the residual $\delta P$. 

Because of its coupling with the outer orbit, this pulsar's intrinsic spin-frequency derivative, $\dot{\nu}_0$, cannot be measured directly as usual for a solitary pulsar or derived from measurements for binary pulsars. Assuming a characteristic age $\tau_c=-\nu/(2\dot\nu)=10^{10}$~yr, a typical value for millisecond pulsars with a period of 3.2~ms, we preliminarily took $\dot{\nu}_0 = -4.95\times10^{-16}$~Hz\,s$^{-1}$  -- a better value will be determined later in the model fit. Ignoring the evolution of $\dot{\nu}_0$ and taking the tertiary mass of 1.0~M$_{\rm odot}$, we linearly set 16 observed spin periods over the MJD range covered by the FAST observations, and performed a four-dimensional grid search over $f_{\rm orb2} \in (1/2,000,000, \, 1/8,000) d^{-1}$, $T_{\rm 02} \in (-P_{\rm orb2}/2, \, P_{\rm orb2}/2)$, $\omega_2 \in (0^\circ, \, 180^\circ)$, and $e_2 \in (0.0, 0.96)$ to minimize $\delta P$, with 1000 to 5000 grids for these parameters. Fig.~\ref{Fig:fit}A shows the fitting results. From the FAST data alone, the eccentricity $e_2$ and the orbital frequency $f_{\rm orb2}$ are coupled and cannot be well determined from the best fit.

\subsection{From the Gaia geometry and outer companion mass}

For the semi-major axis of the relative outer orbit of the tertiary and the mass center of the pulsar and the WD, $a_{\rm out}$, which can be linked to the total mass $M_{\rm t}$ of the system via Kepler's third law,
\begin{equation}
    \frac{a_{\rm out}^3}{P_{\rm orb2}^2}=\frac{G M_{\rm t}}{4\pi^2}.
    \label{eq:S2_3}
\end{equation}
The corresponding orbital inclination of the outer orbit $i_2$ can be deduced from
\begin{equation}
    \sin i_2=\frac{x_2}{a_{\rm out}}\frac{M_{\rm t}}{M_{\rm star}}.
    \label{eq:S2_4}
\end{equation}
Using the best‑fit orbital parameters for different values of $P_{\rm orb2}$ and $e_2$, and adopting $M_{\rm star}=1~M_\odot$ and $S_{\perp}=1.1\times10^4$~lt-s, we obtained the total mass of the triple system, $M_{\rm t}$, which has been constrained above in the range between 2.4 and 3.6~$M_\odot$.
The distance between the barycentre of the inner binary and the outer companion can be expressed by
\begin{equation}
    S_2^2=S_\perp^2+S_{\rm r}^2=S_\perp^2+\left(\frac{x_2(1-e_2^2)\sin(\theta_2+\omega)}{1+e_2\cos\theta_2}\frac{M_{\rm t} }{M_{\rm star}}\right)^2,
    \label{eq:S2_1}
\end{equation}
which is related to the semi-major axis of the relative orbit $a_{\rm out}$ of the outer binary through
\begin{equation}
    S_2=\frac{a_{\rm out}(1-e_2^2)}{1+e_2\cos\theta_2}.
    \label{eq:S2_2}
\end{equation}

From the four-dimensional grid search above for a physically viable set of orbital parameters, we found a solution with $M_{\rm t}$ in the range $2.4$--$3.6~M_\odot$ and $\sin i_2\le 1$.  As shown in Fig.~\ref{Fig:fit}B, we confirm that Gaia DR3 171833395176138240 is the outer companion of PSR~J0435+3233, and that the tertiary orbital period is between 50 and 150\,yr.

\subsection{Integrating the gravitational three-body effects on the inner orbit into TEMPO2}

The inner companion is most likely a WD; in this case, the observed $\dot{f}_{\rm orb1}$ and $\dot{x}_1$ in Table~\ref{tab:FAST-timing-solution} can constrain the orbital geometry. The observed $\dot{f}_{\rm orb1}$ is composed of
\begin{equation}
    \dot{f}_{\rm orb1}=\dot{f}^{\rm D}_{\rm orb1}+\dot{f}^{\rm tD}_{\rm orb1}+\dot{f}^{\rm t}_{\rm orb1}+\dot{f}^{\rm GW}_{\rm orb1},
\end{equation}
where $\dot{f}^{\rm D}_{\rm orb1}$ originates from the classical Doppler effect, $\dot{f}^{\rm tD}_{\rm orb1}$ from {\it the transverse Doppler effect}, $\dot{f}^{\rm t}_{\rm orb1}$ from the perturbation by the outer companion, and $\dot{f}^{\rm GW}_{\rm orb1}$ from gravitational wave emission. The detailed expressions for the three‑body terms $\dot{f}^{\rm D}_{\rm orb1}$, $\dot{f}^{\rm tD}_{\rm orb1}$ and $\dot{f}^{\rm t}_{\rm orb1}$ are given below, while $\dot{f}^{\rm GW}_{\rm orb1}$ is negligibly small. Similarly, the observed $\dot{x}_1$ can be written as
\begin{equation}
    \dot{x}_1=\dot{x}_1^{\rm D}+\dot{x}_1^{\rm tD}+\dot{x}_1^{\rm t},
\end{equation}
where the three contributions are, respectively, the classical Doppler term, the transverse Doppler term, and the perturbation from the outer companion.

If one takes the timing parameters of PSR~J0435+3233 in Table~\ref{tab:FAST-timing-solution} obtained from the FAST data, one can find small additional changes in the orbital period and projected semi-major axis of the inner orbit, exceeding the classic Doppler effect $\dot{f}_{\rm orb1}^{\rm D}$ and $\dot{x}_1^{\rm D}$ \citep{Damour+1992PhRvD..45.1840D} by amounts of 
\begin{equation}
\begin{aligned}
     \dot{f}_{\rm orb1}-\dot{f}_{\rm orb1}^{\rm D}&=\dot{f}_{\rm orb1}-\frac{\dot{\nu}-\dot{\nu}_0}{\nu}f_{\rm orb1}\\&<-1.88(5)\times10^{-22}~{\rm s^{-2}},\\
     \dot{x}_1-\dot{x}_1^{\rm D}&=\dot{x}_1+\frac{\dot{\nu}-\dot{\nu}_0}{\nu}{x}_1\\&\approx4.9\times10^{-13}~\text{lt-s s}^{-1},
\end{aligned}
\end{equation}
where $\dot{\nu}_0$ is the unknown intrinsic spin-frequency derivative, and one can take the assumed value above. The small discrepancies likely originate from gravitational three-body effects in the triple system, which we consider in detail below.

For a triple pulsar, the {\it transverse Doppler effect} is the most significant relativistic contribution to the timing. The delay induced by this effect, $\Delta_{\rm tD}$, is given by the integral expression
\begin{equation}
    \frac{{\rm d}\Delta_{\rm tD}}{{\rm d}t}=\frac{|\vec v_1+\vec v_2|^2}{2}+{\rm constant},
    \label{eq:Dt}
\end{equation}
where $\vec v_1$ is the pulsar's orbital velocity in the inner binary and $\vec v_2$ is the velocity of the inner binary's center of mass in the outer orbit. The contributions from $v_1^2$ and $v_2^2$ are parts of the Einstein delay \citep{Damour+1992PhRvD..45.1840D}. In a triple system, the cross term $\vec v_1\cdot\vec v_2$ makes a unique contribution to the timing, as previously demonstrated in the case of the triple pulsar PSR~J0337+1715 \citep{Ransom+2014Natur.505..520R}. Since $\vec v_2$ varies much more slowly than $\vec v_1$ and the inner orbit is nearly circular, 
Eq.~\eqref{eq:Dt} can be written as
\begin{equation}
    \Delta_{\rm tD}=\int\vec v_1 {\rm d}t \cdot\vec v_2\approx a_1(v_{2,x}\cos\Phi_1+v_{2,y}\sin\Phi_1),
\end{equation}
where $v_{2,x}$ is the component of $\vec v_2$ along the ascending node of the pulsar's inner orbit, $v_{2,y}$ is the component perpendicular to it within the orbital plane, $a_1\equiv x_1/\sin i_1$ is the pulsar's semi-major axis around the inner binary barycentre, and $\Phi_1=2\pi f_{\rm orb1}(t-T_{\rm asc1})$ is the mean anomaly. Combining $\Delta_{\rm tD}$ with the R\"{o}mer delay $\Delta_{\rm R1}$, one get 
\begin{equation}
    \Delta_{\rm R1}+\Delta_{\rm tD} \approx (x_1+a_1v_{2,y})\sin\Phi_1+a_1v_{2,x}\cos\Phi_1.
\end{equation}
One then can find an effective projected semi-major axis $x_1^{\rm eff}$ and an effective mean anomaly $\Phi_1^{\rm eff}$ for the inner orbit:
\begin{equation}
\begin{aligned}
    &x_1^{\rm eff}\approx x_1+a_1v_{2,y},\\
    &\Phi_1^{\rm eff}\approx\Phi_1+\frac{v_{2,x}}{\sin i_1}.
\end{aligned}
\end{equation}
Consequently, the {\it transverse Doppler effect} induces a change in the inner orbital period of
\begin{equation}
    \dot{f}_{\rm orb1}^{\rm tD}\approx \frac{1}{2\pi}\frac{{\rm d}^2({v}_{2,x}/\sin i_1)}{{\rm d} t^2}.
\end{equation}
It contributes to $\dot{x}_1$ by amount of  
\begin{equation}
    \dot{x}_1^{\rm tD}=\frac{{\rm d}({a}_1v_{2,y})}{{\rm d} t}.
    \label{eq:tD}
\end{equation}

\begin{figure}
    \centering
    \includegraphics[width=0.9\linewidth]{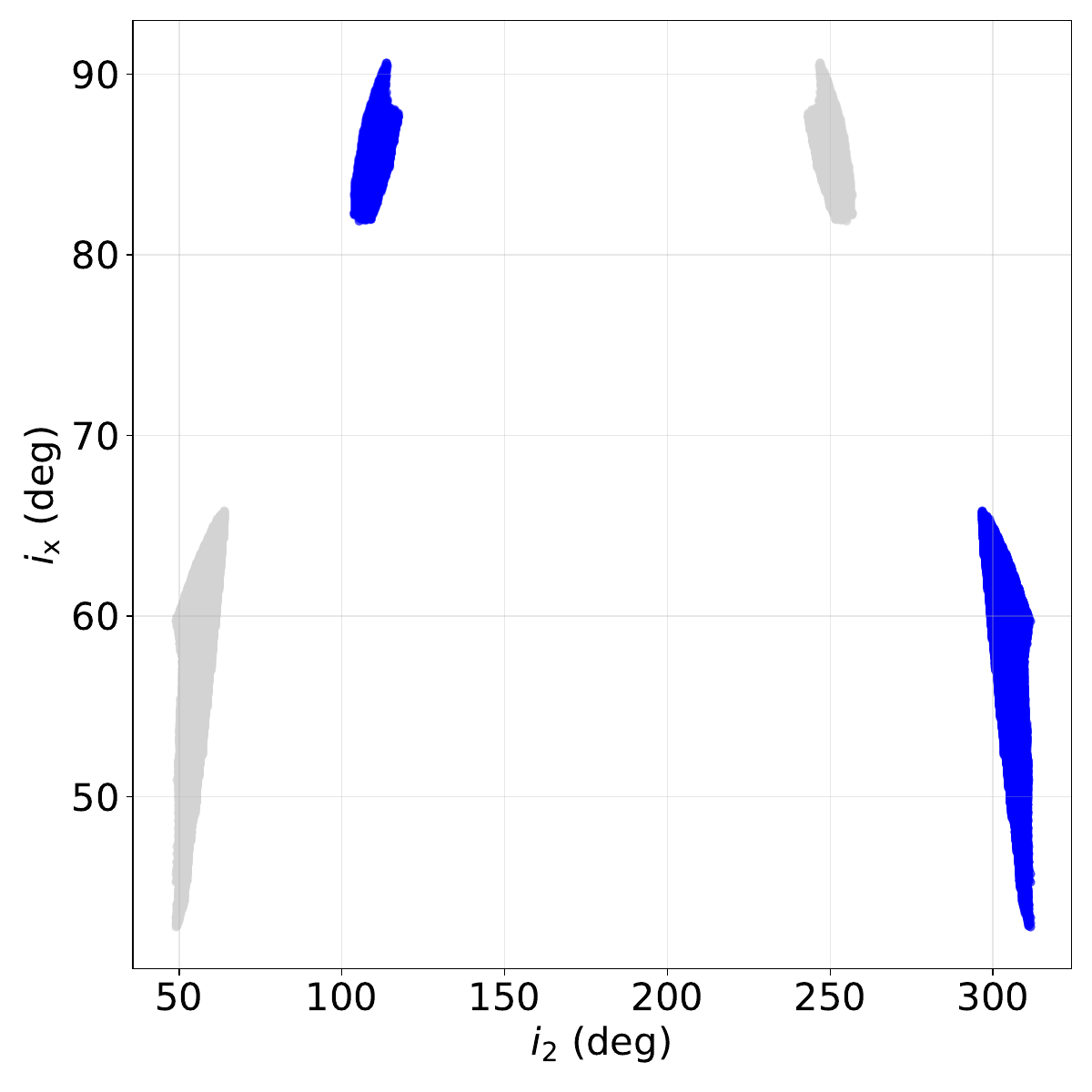}
    \caption{Possible ranges of the orbit inclination angles for the PSR J0435+3233 triple system. To explain the radio and optical/infrared results, the inner and outer orbits are either almost perpendicular or have a moderate mutual inclination of about $i_{\rm x}= 54^\circ$. The grey regions were ruled out by the direction of relative proper motion between the pulsar and the outer companion.
    }
    \label{fig:Imut}
\end{figure}

In addition to the {\it transverse Doppler effect}, the tertiary perturbations cause variations of the inner orbit's parameters, which can be determined using the Lagrangian planetary equations \citep{Valtonen+2006tbp..book.....V}. We considered only the lowest-order term, the quadrupole term, in the perturbing function. For a nearly circular inner orbit,
the time evolution of the inner orbital inclination angle $i_1$ driven by the tertiary perturbations is \citep{Rasio+1994ApJ...427L.107R, Joshi+1997ApJ...479..948J}
\begin{equation}
    \dot{i}_1^{\rm t}=\frac{3\pi}{P_{\rm orb1}}\frac{M_{\rm star}a_{\rm in}^3}{M_{\rm in}S_2^3}{s_{x}s_{z}}.
\end{equation}
We have defined $s_{x}=S_{2,x}/S_2$, $s_{y}=S_{2,y}/S_2$, and $s_{z}=S_{2,z}/S_2$, where $S_{2,x}$, $S_{2,y}$, and $S_{2,z}$ denote the projections of $S_2$ in Eq.~\eqref{eq:S2_2} onto the respective directions. The time evolution of the inner orbit's projected semi-major axis caused by the tertiary perturbation is 
\begin{equation}
    \dot{x}^{\rm t}_1=x_1\cot i_1\,\dot{i}_1.
    \label{eq:perx}
\end{equation}

In addition, the third-body perturbation can also modify $f_{\rm orb1}$. For a nearly circular orbit, $2\pi f_{\rm orb}=\dot{M}+\dot\omega$. According to the Lagrangian planetary equations, $\dot{f}_{\rm orb1}^{\rm t}$ is \citep{Valtonen+2006tbp..book.....V}
\begin{equation}
    \dot{f}_{\rm orb1}^{\rm t}
    =\frac{1}{P_{\rm orb1}}\frac{M_{\rm star}a_{\rm in}^3}{M_{\rm in}}\dot{Q},  
    \label{eq:perf}
\end{equation}
which is an important contribution to the inner orbital frequency derivative from the tertiary. Here, $\dot{Q}$ describes the time-variation of $Q=\left(3s_z^2-\frac{3}{2}\cot i_1 s_ys_z-1\right)\Big/S_2^3$. 
 
These effects should be able to produce the observed $\dot{f}_{\rm orb1}$, $\ddot{f}_{\rm orb1}$, $\dot{x}_1$, and $\ddot{x}_1$, providing further constraints on the orbital geometry of the triple system. 

Following \citet{Rasio+1994ApJ...427L.107R} and \citet{Joshi+1997ApJ...479..948J}, the tertiary perturbation also leads to the inner orbit's periastron advance and eccentricity evolution. According to the Lagrangian planetary equations, they can be written as
\begin{equation}
\begin{aligned}
    \dot{e}_1^{\rm t}&=-e_1\frac{15\pi}{P_{\rm orb1}}\frac{M_{\rm star}a_{\rm in}^3}{M_{\rm in}S_2^3}s_e\sqrt{1-s_z^2-s_e^2},\\
    \dot{\omega}_1^{\rm t}&=\frac{3\pi}{P_{\rm orb1}}\frac{M_{\rm star}a_{\rm in}^3}{M_{\rm in}S_2^3}\left[5s_e^2+s_z^2\right.\\&~~~~~~~~~~~~~~~~~~~~~~~~~~~~\left.-\cot i_1s_ys_z-2\right],
\end{aligned}
\end{equation}
here we have defined $s_e=S_{2,e}/S_2$, and $S_{2,e}$ is the component of $\vec{S}_2$ along the periastron passage of the inner orbit. 
General Relativity (GR) also introduces a periastron advance that can be described by the first-order post-Newtonian (1PN) theory as \citep{Damour+1992PhRvD..45.1840D}
\begin{equation}
\begin{aligned}
    \dot \omega_{\rm GR} &=3\left(\dfrac{P_{\rm orb1}}{2\pi}\right)^{-5/3}(GM_{\rm in})^{2/3}(1-e_1^2)^{-1}\\
&\simeq 1.5 \times 10^{-4} \, \text{rad/yr}
\end{aligned}
\label{eq:omega GR bar}
\end{equation}
$\dot \omega_{\rm GR}$ are included in our timing analysis as slow changes of $\epsilon_1\equiv e_1\sin\omega_1$ and $\epsilon_2\equiv e_1\cos\omega_1$. However, because the inner orbit is very circular ($e_1=0.00016$), the small changes of $\dot{e}_1^{\rm t}$ and $\dot{\omega}_1^{\rm t}$ could be ignored in this current timing analysis.

In the above analysis, the perturbation potential is averaged over the inner orbital period $P_{\rm orb1}$. The tertiary perturbation also causes periodic changes of the inner orbital parameters within each $P_{\rm orb1}$. The magnitude of variations of R\"{o}mer delay can be estimated by $\mathcal{K} = x_1{M_{\rm star}a_{\rm in}^3}/(M_{\rm in}{S_2^3})\sim3\times10^{-8}$~s, and compared to current timing precision, such variations cannot be ignored. The orbital-parameter variations with the inner orbital phase $\Phi_1$ can be derived from the Lagrangian planetary equations and expressed by: 
\begin{equation}
\begin{split}
\frac{\delta a_{\rm in}}{a_{\rm in}} &= \mathcal{K} \left[ 3s_xs_y\sin 2\Phi_1 + \frac{3}{2}(s_x^2 - s_y^2) \cos 2\Phi_1 \right], \\
\delta i_1 &= \frac{3}{4} \mathcal{K} \, s_z \left( s_x \sin 2\Phi_1 - s_y \cos 2\Phi_1 \right), \\
\delta \Phi_1 &= \mathcal{K} \left[ \left( \frac{21}{4}s_xs_y+ \frac{3}{4} s_z s_x \cot i_1 \right) \cos 2\Phi_1 \right. \\
&\qquad \left. + \left( -\frac{21}{8}(s_x^2 - s_y^2) + \frac{3}{4} s_z s_y \cot i_1 \right) \sin 2\Phi_1 \right], \\
\delta \epsilon_1 &= \mathcal{K} \left[ \left( 1 - \frac{15}{4}s_x^2 + \frac{3}{4}s_y^2 \right) \sin\Phi_1 + \frac{9}{2} s_xs_y\cos\Phi_1 \right. \\
&\qquad \left. + \frac{1}{4}(s_x^2 - s_y^2) \sin 3\Phi_1 - \frac{1}{2} s_xs_y\cos 3\Phi_1 \right], \\
\delta \epsilon_2 &= \mathcal{K} \left[ \left( 1 + \frac{3}{4}s_x^2 - \frac{15}{4}s_y^2 \right) \cos\Phi_1 + \frac{9}{2} s_xs_y\sin\Phi_1 \right. \\
&\qquad \left. + \frac{1}{4}(s_x^2 - s_y^2) \cos 3\Phi_1 + \frac{1}{2} s_xs_y\sin 3\Phi_1 \right].
\end{split}
\label{eq:fast_per1}
\end{equation}
Their contribution to the R\"{o}mer delay of the inner orbit $\Delta_{\rm R1}$ can be approximated as:
\begin{equation}
\begin{aligned}
    \delta \Delta R_1 & =x_1 \cos\Phi_1 \, \delta\Phi_1 + \delta x_1 \sin\Phi_1 \\&- \frac{1}{2} x_1 \, \delta \epsilon_1 \left( \cos 2\Phi_1 + 3 \right) + \frac{1}{2} x_1 \, \delta \epsilon_2 \sin 2\Phi_1.
\end{aligned}
\label{eq:fast_per2}
\end{equation}
Combining Eqs.~\ref{eq:fast_per1} and~\ref{eq:fast_per2}, we obtain
\begin{equation}
\begin{aligned}
\delta\Delta_{\rm R1}
=~&\mathcal{K}x_1\left[-1+\frac{43}{16}s_x^2+\frac{5}{16}s_y^2+\frac{3}{4}s_y s_z\cot i_1\right]\sin\Phi_1\\
&+\mathcal{K}x_1\left[-\frac{19}{8}s_xs_y+\frac{3}{4}s_xs_z\cot i_1\right]\cos\Phi_1\\
&+\mathcal{K}x_1\frac{3}{16}(s_x^2-s_y^2)\sin3\Phi_1\\
&-\mathcal{K}x_1\frac{3}{8}s_xs_y\cos3\Phi_1.
\end{aligned}
\label{eq:combined_fast_per}
\end{equation}
The tertiary perturbation leads to systematic changes up to the $3\Phi_1$ term, affecting the Shapiro delay measurements for the first three harmonics up to $h_3$, but not the fourth harmonic $h_4$.

\begin{figure*}[t]
    \centering
    \includegraphics[width=0.9\linewidth]{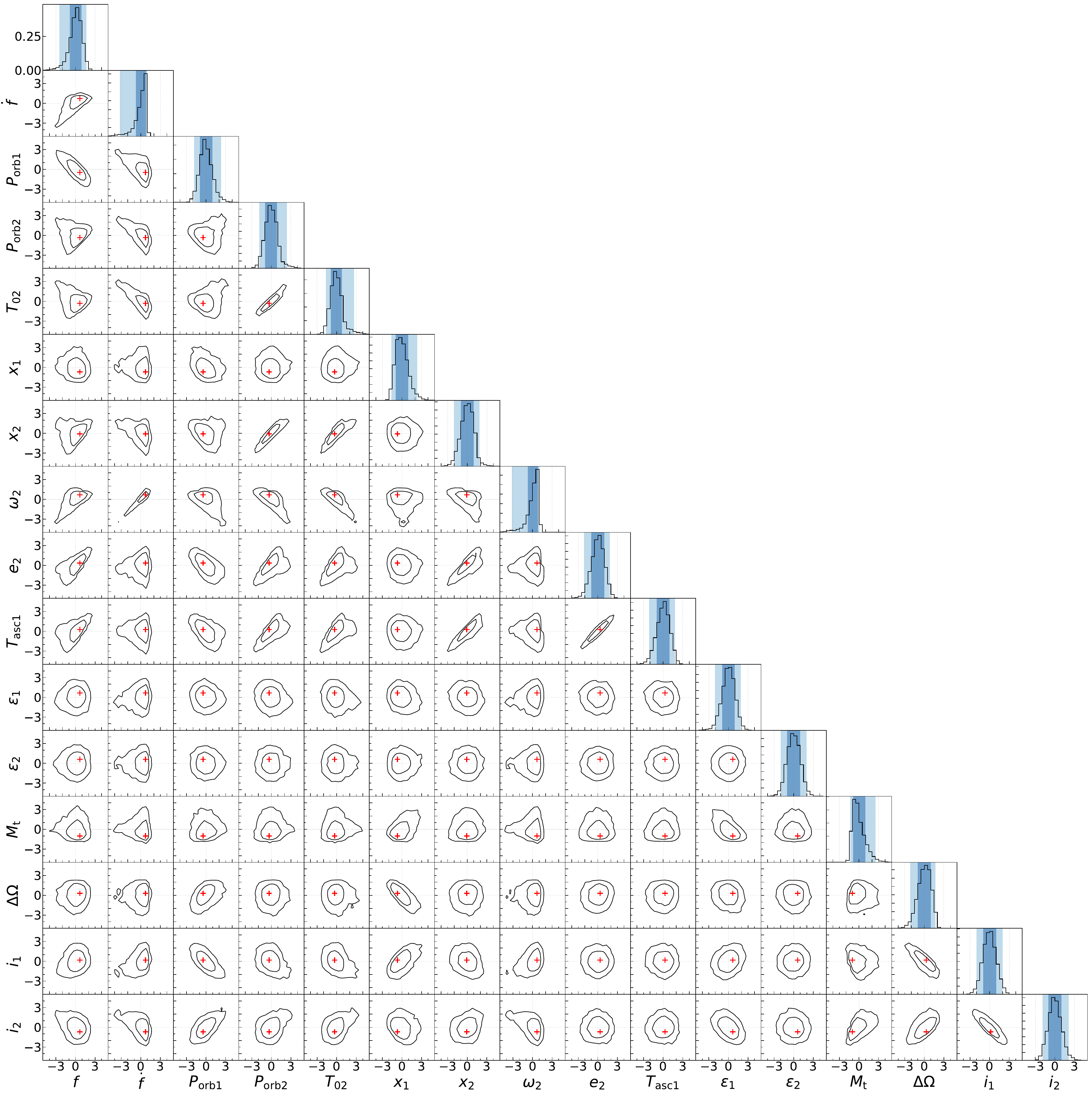}
   \caption{The probability distributions (corner plots) of the fitted parameters in the timing solution A in Table~\ref{tab:timing_solution_final} (similar for solution B, not shown) using \textsc{tempo2} \citep{Hobbs+2006MNRAS.369..655H}. They are calculated using the nested sampling technique based on the \texttt{ReactiveNestedCalibrator} function in \textsc{UltraNest} \citep{Higson+2019MNRAS.483.2044H, Buchner+2021JOSS....6.3001B}. All parameters have been scaled by their mean and standard deviation, and all subpanels are plotted over the range $\pm5\sigma$ with two contour lines for the 68\% and 95\% possibilities. The red cross marks the maximum probability. The dark shadow in the histogram marks the $\pm1\sigma$ range, and the shallow shadow marks the $\pm2\sigma$ range.
    }
    \label{fig:posterior}
\end{figure*}

Based on the three-body gravitational effects discussed above, we developed a new timing model in \textsc{tempo2} \citep{Hobbs+2006MNRAS.369..655H} that employs the Keplerian orbital parameters of both orbits, the two orbital inclinations, the mutual inclination, and the total system mass to predict TOAs. The pulsar mass $M_{\rm PSR}$, WD mass $M_{\rm WD}$, and the outer companion mass $M_{\rm star}$ are derived from these parameters using Eq.~\eqref{eq:mf_in} and \eqref{eq:S2_3}.
The new timing formula is now expressed as
\begin{equation}
\begin{aligned}
t_{\rm b}=t+\left(1+\frac{{\rm d}\Delta_{\rm R2}}{{\rm d}t}\right)\Delta_{\rm R1}(t)+\Delta_{\rm R2}(t)\\
+\Delta_{\rm S1}(t)+\Delta_{\rm T}(t)+\Delta_{\rm E2}(t),
\label{eq:timing_formula}
\end{aligned}
\end{equation}
where $t_{\rm b}$ is the barycentric arrival time, $\Delta_{\rm R2}$ is the R\"{o}mer delay of the outer orbit, respectively, and $\Delta_{\rm S1}$ is the Shapiro delay of the inner binary, $\Delta_{\rm E2}$ is the Einstein delay of the outer orbit. The Shapiro delay and Einstein delay are calculated following \citet{Damour+1992PhRvD..45.1840D}, assuming that general relativity is correct. The orbital parameters of the inner binary evolve under the perturbation of the outer companion, as given by Eqs.~\eqref{eq:perx},~\eqref{eq:perf}, and~\eqref{eq:combined_fast_per}. 
The orbital parameters of the outer orbit are evolving slowly, which can be calculated by the numerical integration using the Lagrangian planetary equation with the perturbing function averaged over the inner orbital period.

\begin{table*}[t]
\centering
    \caption{Timing parameters from fitting data with the modified \textsc{tempo2} model and key derived quantities for PSR~J0435+3233 (two possible solutions A and B). The reported values are the mean values of the posterior distributions, and numbers in parentheses are the standard deviations expressed in the magnitude of the last digits. We adopt the uncertainties of some fixed parameters (astrometric parameters and dispersion measure) in the nested sampling analysis from the \textsc{tempo2} fit to FAST data. We took the prior ranges of  $S_{\perp}=1.1(2)\times10^4$ lt-s, $\dot{\nu}<0$, and $M_{\rm PSR}=1.1$ to 2.2$~M_\odot$ for our analysis.
    } 
    \label{tab:timing_solution_final}
    \scriptsize  
    \renewcommand{\arraystretch}{0.85}
    \setlength{\tabcolsep}{1pt}
\noindent 
\begin{tabular}{lcc} 
\hline
 	   \multicolumn{3}{c}{General information}  \\     \hline                          
FAST TOAs MJD range      &        \multicolumn{2}{c}{59101 to 60877}  \\
Number of FAST TOAs      &        \multicolumn{2}{c}{192}       \\
EFAC of FAST TOAs        &        \multicolumn{2}{c}{0.97}    \\
Fermi TOAs MJD range      &       \multicolumn{2}{c}{54887 to 61001}  \\
Number of Fermi TOAs      &       \multicolumn{2}{c}{16}         \\
EFAC of Fermi TOAs        &       \multicolumn{2}{c}{1.53}     \\
Solar System ephemeris    &       \multicolumn{2}{c}{DE440}       \\
Reference epoch (MJD)    &        \multicolumn{2}{c}{59997.5}      \\
\hline\\[-3mm]
		Parameter    & Solution A & Solution B \\
\hline\\[-3mm]
 	   \multicolumn{3}{c}{Basic pulsar parameters}    \\  
\hline 
Right ascension
, $\alpha$ (J2000)   &      $04^{\rm h}35^{\rm m}33^{\rm s}.7608626(28)$ & $04^{\rm h}35^{\rm m}33^{\rm s}.7608612(28)$ \\
Declination
, $\delta$ (J2000)       &      $+32^\circ33'07''.93347(17)$ & $+32^\circ33'07''.93354(17)$\\
Proper motion in RA
\;  (mas\,yr$^{-1}$) & 1.28(3) & 1.27(3) \\
Proper motion in DEC
\;  (mas\,yr$^{-1}$) & -3.29(15) & -3.33(15) \\
Spin frequency, $\nu$ (Hz) &  312.7111521(5) & 312.7111514(8) \\  
Spin frequency derivative, $\dot{\nu}$ (Hz $\rm s^{-1}$) & $-8(8)\times10^{-16}$ & $2.6(16)\times10^{-15}$ \\
\hline\\[-3mm]
 	\multicolumn{3}{c}{Inner Keplerian parameters for pulsar orbit}   \\  
\hline 
Orbital period, $P_{\rm orb1}$ (day) &  7.998147072(15) & 7.998147028(22) \\ 
Projected semi-major axis, $x_1$ (lt-s) & 7.9781313(4) &  7.9780676(22) \\ 
Time of ascending node passage, $T_{\rm asc1}$ (MJD) & 59997.508015(8) & 59997.508023(9)\\
First Laplace parameter, $\epsilon_1=e_1\sin\omega_1$   & 0.000139068(16) & 0.000139073(21)  \\ 
Second Laplace parameter, $\epsilon_2=e_1\cos\omega_1$ & -0.000084130(12) & -0.000084130(13) \\ 
Inner orbital inclination, $i_1$ (deg) & $107.7(6)$ &  111.0(14) \\
\hline\\[-3mm]
    \multicolumn{3}{c}{Outer Keplerian parameters for inner binary barycenter}   \\
\hline 
Orbital period, $P_{\rm orb2}$ (day) & 26850(10) & 26872(19) \\
Projected semi-major axis, $x_2$ (lt-s) & 2502.7(10) & 2504.3(16) \\ 
Time of periastron passage, $T_{\rm 02}$ (MJD) & 64802.8(4) & 64803.5(7)\\
Longitude of periastron passage, $\omega_2$ (deg) & 31.868(26) & 31.81(5) \\ 
Orbital eccentricity, $e_2$ &  0.59822(6)  & 0.59825(6) \\ 
Outer orbital inclination, $i_2$ (deg) & 146.3(11) &  150.0(6) \\ 
\hline\\[-3mm]
   \multicolumn{3}{c}{Other fitting parameters}   \\
\hline
System total mass, $M_{\rm t}$ ($M_\odot$) & 2.35(10) & 2.78(23) \\
Relative ascending-node longitude, $\Delta\Omega_{\rm asc}$ (deg) & 108.1(12) & 307.5(11) \\
\hline\\[-3mm]
	   \multicolumn{3}{c}{Derived parameters} \\    
\hline 
Galactic longitude, $l$ (deg)         &       	\multicolumn{2}{c}{168.15524} \\
Galactic latitude, $b$ (deg)          &         \multicolumn{2}{c}{-10.00888} \\
Pulsar spin period, $P$ (s)  &  0.003197839256(5) & 0.003197839282(8) \\  
Pulsar mass, $M_{\rm PSR}$ ($M_\odot$) & 1.16(5) & 1.34(15) \\
Inner companion mass, $M_{\rm WD}$ ($M_\odot$)  &  0.273(8) & 0.305(23)  \\
Inner binary total mass, $M_{\rm in}$ ($M_\odot$) & 1.44(6) & 1.65(17) \\
Outer companion mass, $M_{\rm star}$ ($M_\odot$) & 0.91(5) & 1.13(7) \\ 
Mutual orbit inclination angle, $i_{\rm X}$ (deg) & 84.9(8) & 53.5(17) \\
Transverse distance between star and pulsar, $S_{\perp}$ (lt-s) &  $1.190(19)\times10^4$ & 1.273(35)$\times10^{4}$\\
Semi-major axis of the inner orbit, $a_{\rm in}$ (lt-s) & 44.1(6) & 46.1(16) \\
Eccentricity of the inner orbit, $e_{\rm in}$ & 0.000162535(15) & 0.000162540(19)\\
Semi-major axis of the outer orbit, $a_{\rm out}$ (lt-s) & $1.164(16)\times10^4$ & $1.23(3)\times10^{4}$ \\
Semi-major axis of pulsar from inner binary barycenter, $a_1$ (lt-s) & 8.376(27) & 8.55(8) \\
Pulsar periastron longitude around inner binary barycenter, $\omega_1$ (deg)  & 121.172(5) &121.171(5) \\
Pulsar ascending node longitude around inner binary barycenter, $\Omega_1$ (deg) & 105 & 264 \\
Semi-major axis of inner barycenter from triple barycenter, $a_2$ (lt-s) & $4.52(13)\times10^3$ & 5.01(9)$\times10^3$ \\
 Ascending node longitude of inner barycenter around triple barycenter, $\Omega_2$ (deg) & 213 & 212\\
		\hline
    \end{tabular}
\end{table*}

\begin{figure}
    \centering
    \includegraphics[width=0.95\linewidth]{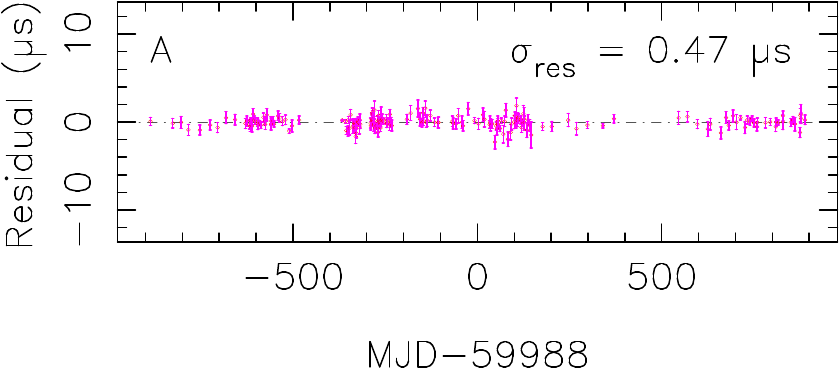}
    \includegraphics[width=0.95\linewidth]{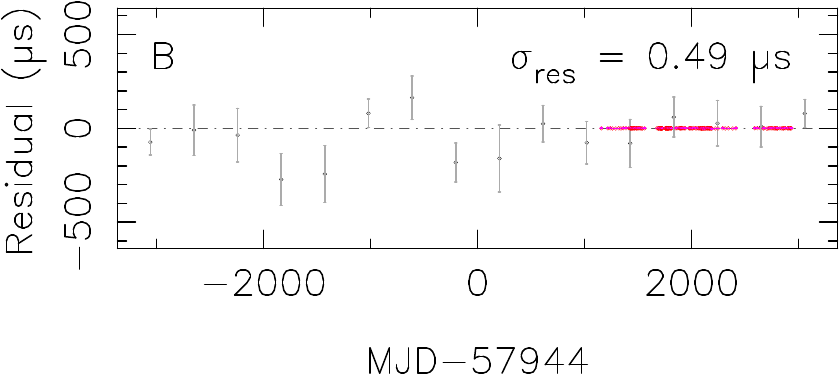} 
    \caption{Timing residuals of PSR J0435+3233 based on the timing solution A. 
    Panel A shows the timing residuals from FAST TOAs only, while panel B shows the timing residuals from Fermi TOAs (gray) and FAST TOAs. The timing residuals of solution B are almost the same. 
    }
    \label{fig:res_final}
\end{figure}

\begin{figure}[t]
    \centering
    \includegraphics[width=0.98\linewidth]{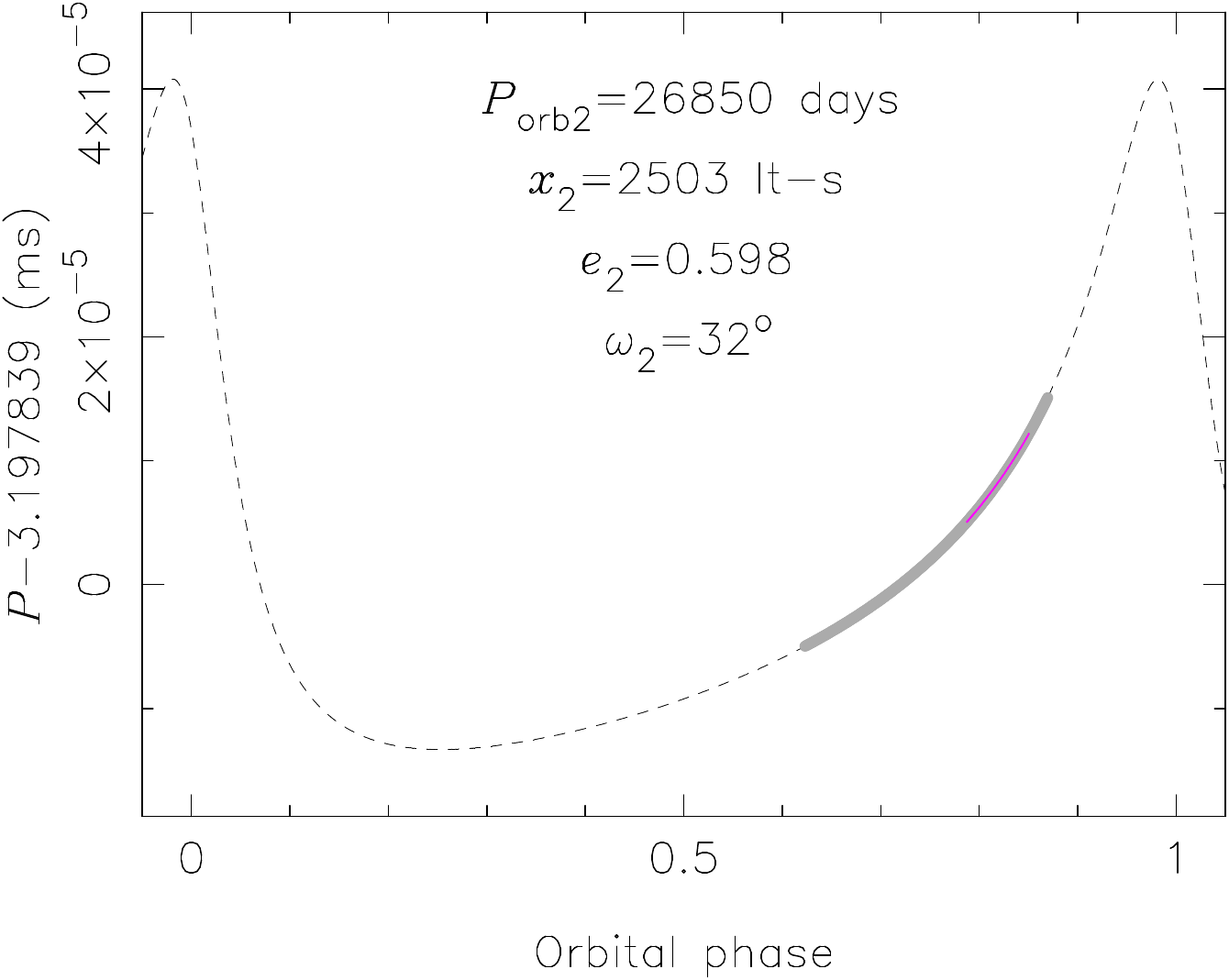}
    \caption{Predicted pulsar spin-period variations along the outer orbit based on the timing solution A.  The orbital phase is defined as the mean anomaly measured from $\omega_2$. The FAST observational time span occupies only a small fraction of the outer orbit, indicated by the magenta line. The Fermi data span is indicated by the thick grey line.
    }
    \label{fig:P0-phase}
\end{figure}

\subsection{Searching for viable geometry parameters of the triple system}

Once the Keplerian orbital parameters of the two orbits are known, for a given $M_{\rm star}$ and $S_{\rm \perp}$, the inclination angle of the outer orbit, $i_2$, and the total mass of the inner binary, $M_{\rm in}$, can be derived using Eqs.~\eqref{eq:S2_1}, \eqref{eq:S2_2}, \eqref{eq:S2_3}, and \eqref{eq:S2_4}. Given the newly fitted Shapiro delay parameter $h_4$ of the pulsar caused by the WD (see Table~\ref{tab:FAST-timing-solution}), one can also determine the inclination angle of the inner orbit $i_1$ using Eq.~\eqref{eq:Shapiro}. Then, for a given relative angle between the orbits' ascending nodes, $\Delta\Omega=\Omega_2-\Omega_1$, one can determine the relative orbital orientation. The secular evolution of the inner orbit's parameters can be determined using Eqs.~\eqref{eq:tD}, \eqref{eq:perf} and \eqref{eq:perx}. 

Using the parameters from the grid search shown in Fig.~\ref{Fig:fit} in the area for  $\delta P < 2 \delta P_{\rm min}$ and adopting $M_{\rm star}$ in the range 0.62 to 1.34\,$M_\odot$ with steps of 0.005~$M_\odot$, $S_{\perp}$ between 7000 and 15000~lt-s with steps of 100~lt-s (the mass estimate is less reliable than the distance estimate), $h_4$ between $\pm 2\sigma_{\rm h_4}$ with steps of 0.5 $\sigma_{\rm h_4}$, and a pulsar mass between 1.1 and 2.2~$M_\odot$ \citep{Alsing+2018MNRAS.478.1377A, Muller+2025PhRvL.134g1403M}, we searched the value ranges of viable geometry parameters and also the full range of the relative ascending‑node longitude $\Delta\Omega \equiv \Omega_2-\Omega_1$ from 0 to $2\pi$ with 1000 steps, and evaluated the combined contributions of the different terms to $\dot{f}_{\rm orb1}$ and $\dot{x}_1$. A geometry is viable only when the predicted $\dot{x}_1$, $\ddot{x}_1$, $\dot{f}_{\rm orb1}$, and $\ddot{f}_{\rm orb1}$ simultaneously match the FAST TOA measurements within their $2\sigma$ uncertainties. We found two possible geometries: the inner and outer orbits are almost perpendicular, or they have a moderate mutual inclination of about 55$^\circ$ (see Fig.~\ref{fig:Imut}). In fact, both $\Delta\Omega$ and $360^\circ-\Delta\Omega$ are possible solutions, corresponding to $(i_1,~i_2)$ and $(180^\circ-i_1,~180^\circ-i_2)$, respectively. However, the relative proper motion in Table~\ref{tab:FAST-timing-solution} and~\ref{tab:optical} excludes the possibility of $i_2>90^\circ$, as pointed out by \citet{Freire2026}. Note also that the geometries are different from the coplanar orbits of the PSR~J0337+1715 triple system  \citep{Ransom+2014Natur.505..520R}. Among all viable orbital geometries, the minimum $\delta P$ is $4.5\times10^{-12}$~ms, corresponding to Keplerian orbital parameters of $P_{\rm orb2}=$73.5~yr, $e_2=0.597$, $x_2=2499$~lt-s, $T_{02}\simeq$MJD~64812.8, and $\omega_2=32^\circ$. 

\subsection{Refining the timing solution of the triple system from TOAs from Fermi and FAST data}

The Fermi data span MJD 54684 to 61205, a much longer time baseline than the FAST data. We can use both FAST and Fermi data to refine the triple-system timing solution. Using the best-fit Keplerian orbital parameters, orbital inclinations, and total system mass above, with small adjustments to the orbital parameters through the modified \textsc{tempo2} fits for the triple system, we obtained a new timing solution for PSR J0435+3233 from the FAST TOA data. With this timing solution, we folded all available Fermi data and found that the pulsar's $\gamma$-ray emission aligns well over almost the entire Fermi data span.

Next, we extracted 16 TOAs of PSR J0435+3233 from the Fermi data following \citet{Ray+2011ApJS..194...17R}, which can provide constraints on the long-term variations caused by the outer orbit. We combined these TOAs with FAST TOAs to form a new set of TOAs. The coordinates, proper motion, and dispersion measure parameters were adopted from the previous best \textsc{tempo2}-fit, and then all TOA values were converted to the solar barycenter, which are used in the following nested sampling analysis. We then refined the model parameters using \textsc{tempo2} \citep{Hobbs+2006MNRAS.369..655H} with the nested sampling technique based on the \texttt{ReactiveNestedCalibrator} function in \textsc{UltraNest} \citep{Higson+2019MNRAS.483.2044H, Buchner+2021JOSS....6.3001B}, with 4000 ``live points'' and a ``log-evidence error'' threshold of 0.1. The prior probability distribution for the transverse distance between the tertiary star and pulsar is taken in the range of  $S_{\perp}=1.1(2)\times10^{4}$~lt-s, the pulsar mass in the range of $M_{\rm PSR}=1.1$--$2.2~M_\odot$ \citep{Alsing+2018MNRAS.478.1377A, Muller+2025PhRvL.134g1403M}, and the spin frequency derivative of $\dot\nu<0$.
We found two sets of timing solutions, named solution A and solution B, as presented in Table~\ref{tab:timing_solution_final}. 
Fig.~\ref{fig:posterior} shows the probability distributions of the fitted parameters from the nested sampling. 
Fig.~\ref{Fig:geo} shows the projected orbits of the pulsar and its companions along our line of sight based on solution A.
We fold the $\gamma$-ray data with the two suites of timing solutions, achieving similar H-test values of  221 (see Fig.~\ref{fig:gammaFoldplot1}).
The timing residuals after fitting the data with the model parameters at the maximum-probability point of the solution A (cf. Fig.~\ref{fig:res}) are shown in Fig.~\ref{fig:res_final}. The relative proper motion predicted by the geometries is about $-0.7$~mas\,yr$^{-1}$ in RA and $-0.5$~mas\,yr$^{-1}$ in DEC, which is consistent with the observational results at the 2$\sigma$ level. Fig.~\ref{fig:P0-phase} shows the predicted pulsar period variations along the outer orbital phase.

\section{Discussions} 

Based on the optical/infrared and $\gamma$-ray archived data, we identified PSR~J0435+3233 as a $\gamma$-ray pulsar in a hierarchical triple system, with a He WD as a close inner binary companion and a Sun-like star as the distant tertiary. We determined the tertiary's outer elliptical orbit, with a period $P_{\rm orb2}=73.5~$yr and an eccentricity $e_2 = 0.598$. This is a unique triple system for detailed multi-band observations and for studying the evolutionary path and dynamic processes of a primordial triple star system.

\begin{figure}
    \centering
    \includegraphics[width=0.8\linewidth]{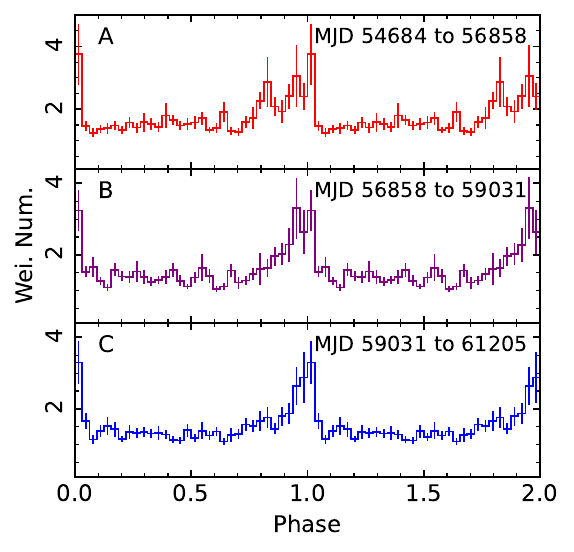}
    \includegraphics[width=0.75\linewidth]{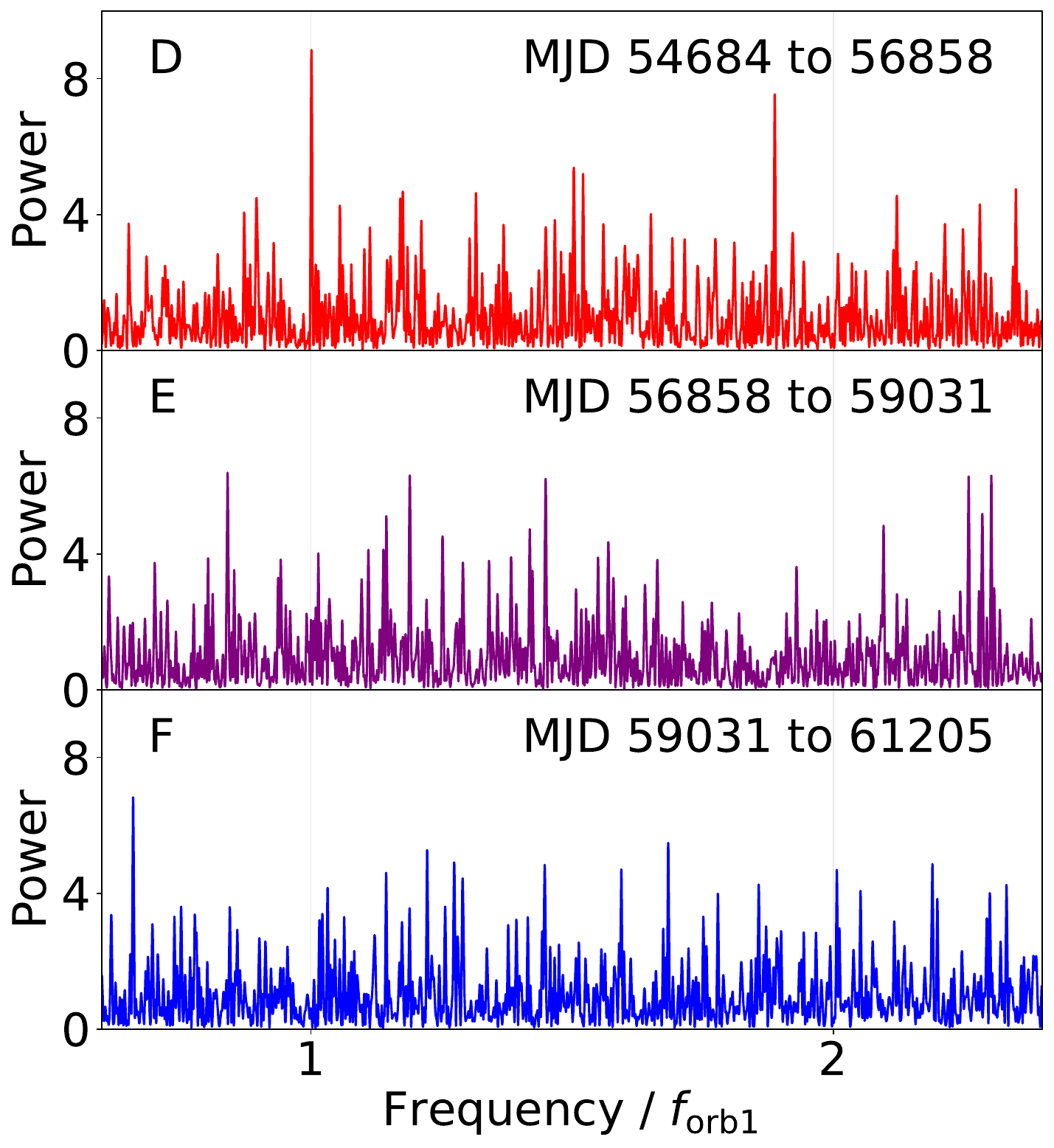}
    \caption{Three upper subpanels (A-C): Folded Fermi $\gamma$-ray profiles in the three MJD ranges with error bars for the $1\sigma$ uncertainty: (A) for MJD $\in(54684, 56858)$, (B) for MJD $\in(56858, 59031)$, and (C) for MJD $\in(59031, 61205)$. Compared with the profile in subpanel (C) with embedded FAST observation duration, the earlier MJD profiles are wider, implying uncorrected periodic perturbations from an imperfect ephemeris of this triple system. 
%
    Three lower subpanels (D-F): 
Temporal power spectra for the pulsar spin phases data of Fermi $\gamma$-ray photons in the three MJD ranges. Uncorrected perturbations of the inner orbit from the tertiary could induce some power on the 
exact frequency for the inner orbital period, as we do see in subpanel (D) for the early MJD range. 
}
    \label{fig:amp}
\end{figure}


\subsection{Residual perturbation from the imperfect timing analysis for the triple system}

Because the tertiary orbit has limited phase coverage in the FAST and Fermi observations, it is hard to constrain the long- and short-term perturbations on the inner orbit from the tertiary. In the long term, the proper motion of the pulsar leads to variations of $x_1$, $P_{\rm orb1}$, $x_2$, and $\omega_2$ through changes in line-of-sight direction \citep{Kopeikin+1996ApJ...467L..93K}, and the motion in the sky plane could not be approximated by a straight line, as pointed out by \citet{Freire2026}. Fortunately, the deviation in TOAs caused by the changes in $x_1$, $P_{\rm orb1}$, and proper motion is about $10$ ns in the FAST observational time span, smaller than the timing residuals. The TOA deviations caused by the changes in $x_2$ and $\omega_2$ can be absorbed by small adjustments of other parameters. Therefore, ignoring these effects currently has little influence on the accuracy of the timing model. 
We ignored these effects in the above fittings.

From the current timing solutions based on limited FAST and Fermi data, errors in the predicted inner orbit induce timing residuals. In particular, in the folded Fermi $\gamma$-ray data, photon pulse phases should change periodically at frequency harmonics of the inner orbital period. Because of limited photons, the integration time for each Fermi profile at a TOA is much longer than the inner-orbit period, so we cannot observe such residuals. However, changes in the long-term integrated profiles and the power at harmonics of the inner orbital period could be used to examine such imperfect timing solutions.
As shown in Fig.~\ref{fig:amp}, we folded the Fermi $\gamma$-ray profiles in the three MJD ranges, and found that the profile in the range with embedded FAST observation duration is narrow, while the profiles in the earlier MJD durations are widened. We also analyzed the temporal power spectra of the pulsar spin-phase data from Fermi $\gamma$-ray photons and found power at the exact frequency of the inner orbital period in the Fermi phase data from the early MJD range, implying that periodic perturbations in this triple system remain uncorrected. 
%
%

Despite imperfect timing from data in the limited tertiary orbital phase range, our analysis using the new timing model of the triple system still achieves high timing precision, yielding a Fermi integrated pulse profile with a high H-test value, which forms the basis for future studies of this system.

\subsection{The Zeipel-Lidov-Kozai effect}

Concurrently, because the tertiary companion ($M_{\rm star}$) orbits the center of mass of the inner subsystem on a wider and highly inclined orbit ($\sim 85^\circ$ or $54^\circ$), the inner binary might be subjected to von Zeipel-Lidov-Kozai (ZLK) eccentricity oscillations \citep{Naoz+2016ARAA..54..441N}. The characteristic timescale of this secular process is 
\begin{equation}
\begin{aligned}
      t_{\rm ZLK} =&\frac{1}{n_\mathrm{in}}\frac{M_{\rm in}}{M_{\rm star}}\bigg(\frac{a_\mathrm{out}\sqrt{1-e^2_\mathrm{2}}}{a_\mathrm{in}}\bigg)^3\\
\simeq & 5 \times 10^4\,{\rm yr} \, \left( \frac{M_{\rm in}}{1.44M_\odot} \right)^{1/2} \left( \frac{M_{\rm star}}{0.91M_\odot} \right)^{-1} \\  &
\left( \frac{a_\mathrm{in}}{44.1\,{\rm lt-s}} \right)^{3/2} \left( \frac{a_\mathrm{out}\sqrt{1-e_\mathrm{2}^2}}{9327\,{\rm lt-s}} \right)^3\nonumber . 
\end{aligned}
\end{equation}
Under the updated system parameters, the timescale for GR pericentre precession ($\sim 1/ \dot{\omega}_{\rm GR} \sim 7 \times 10^3\,\text{yr}$) is remarkably short compared to the ZLK timescale ($t_{\rm ZLK} \sim 5 \times 10^4\,\text{yr}$). Consequently, the fast GR-induced precession overwhelmingly dominates the short-range dynamical force, completely quenching the ZLK eccentricity excitation across all mutual inclinations $I_0$. Numerical integrations of the double-averaged secular equations of motion confirm that the inner orbit remains strictly near-circular ($e_{\rm 1} \approx 0$) throughout its long-term evolution \citep{Liu+2015MNRAS.447..747L}, experiencing no significant eccentricity growth regardless of the orbital alignment.

Given the highly inclined configuration observed today, the external tertiary was likely far from a passive spectator in the system's past; rather, it is highly probable that the tertiary played an active and important role in the early evolutionary history, particularly during the formation phase of the millisecond pulsar binary \citep{2007ApJ...669.1298F, 2016ApJ...822L..24N, 2016ComAC...3....6T}. Before the system settled into its current compact state, the primordial inner binary likely had a wider orbit, where secular gravitational perturbations from the misaligned tertiary could easily dominate short-range forces such as tidal dissipation and post-Newtonian effects. Under these conditions, large-amplitude ZLK oscillations would be efficiently excited, periodically driving the inner binary to extreme eccentricities and dramatically suppressing the periastron distance. Such gravitationally induced orbital shrinkage would naturally facilitate the onset of stable mass transfer or Roche-lobe overflow, thereby assisting the recycling process that ultimately formed the millisecond pulsar-WD binary within this hierarchical framework.

\subsection{Evolution of the PSR J0435+3233 triple system} 

The PSR J0435+3233 triple system is located in the Galactic field, not in a dense cluster, so it is very probably evolved from the primordial triple system, i.e., of the three main-sequence (MS) stars born together, with masses of, for example, $20\, M_\odot$ for the pulsar progenitor, $2\, M_\odot$ for the WD progenitor, and $1\, M_\odot$ for the third star.

The formation pathway of PSR J0435+3233 is similar to the scenario proposed for PSR J0337+1715 \citep{Tauris+2014ApJ...781L..13T}. In this scenario, the inner binary initially consists of two stars with a relatively large mass ratio, leading to unstable mass transfer and a common-envelope phase. After the common envelope is ejected, a He star and an MS star remain in the inner binary. The He star subsequently undergoes a SN explosion and forms an NS, which later accretes matter from the companion star and evolves into a MSP with a He WD companion. Different from the PSR J0337+1715 formation scenario, the tertiary star has a sufficiently wide orbit and does not interact with the inner binary during the mass-transfer and recycling phases.

\begin{figure}
    \centering
    \includegraphics[width=0.9\linewidth]{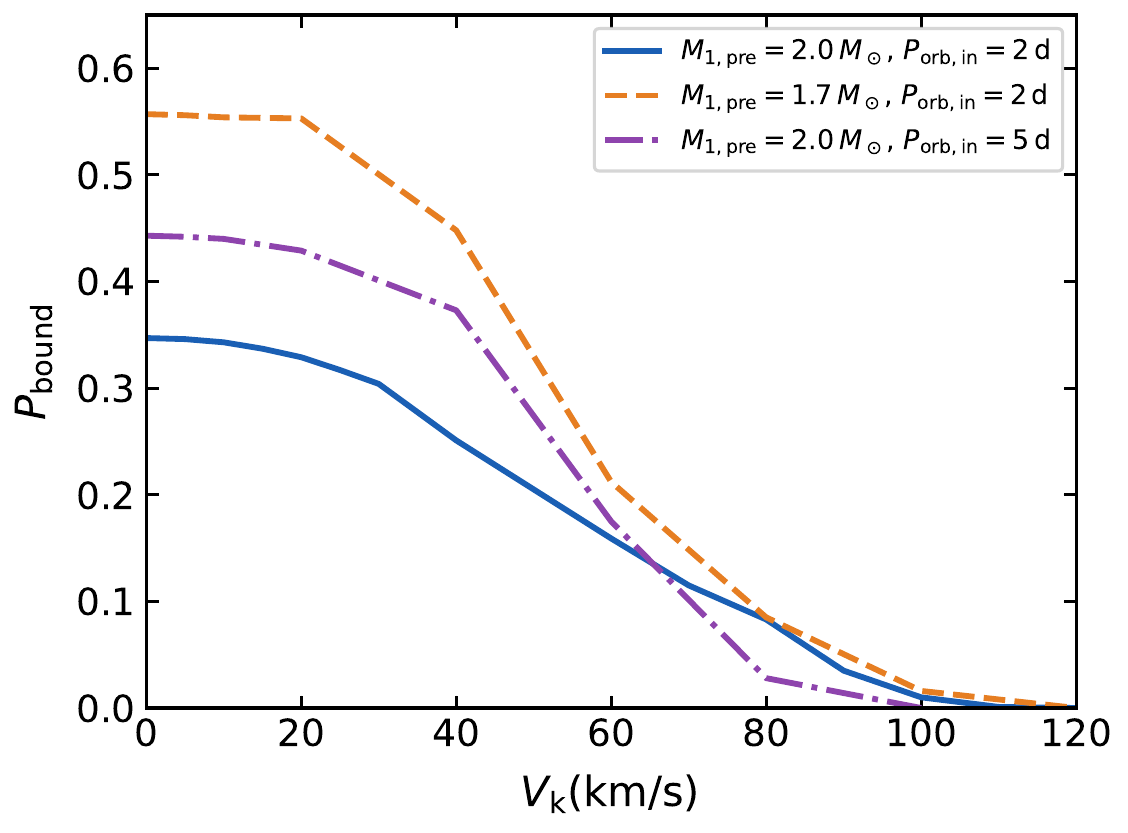}
    \caption{Survival probability of the triple system as a function of the natal kick velocity ($V_{\rm k}$) imparted to the newborn NS. The basic model is plotted by solid line for pre-supernova inner binary with a He star with $M_{1, \rm pre}=2.0\, M_\odot$ at the pre-supernova stage, and a main sequence companion star with a mass of $2.0\, M_\odot$ in an initial orbital period ($P_{\rm orb, in}$) of 2 days, while the tertiary star mass has a mass of $1.3\, M_\odot$ and the outer orbit has a period of 30 yr. We set the newborn NS mass to $1.4\, M_\odot$. The  dashed line is plotted for the alternative case with $M_{1, \rm pre}=1.7\,M_\odot$, and the  dash-dotted line for the case with $P_{\rm orb, in}=5$\,d.
}
    \label{Fig:kick_bound_rate}
\end{figure}

We investigated how supernova explosions in the inner binary, through sudden mass loss and natal kick, determine the survival probability of systems like PSR J0435+3233. Fig.~\ref{Fig:kick_bound_rate} shows the survival probability of the triple system as a function of the natal kick velocity ($V_{\rm k}$) imparted to the newborn NS,
where $V_{\rm k}$ is varied from 0 to 120\,km/s. Our simulations show that a triple system resembling PSR J0435+3233 can survive the supernova explosion provided that the $V_{\rm k}$ does not exceed $\sim 80$\,km/s,
corresponding to the recoil velocity immediately imparted to the inner binary ($w_{12}) \lesssim 40$\,km/s.
We found that even in the limiting case of a symmetric explosion ($V_{\rm k}=0$), the survival probability remains relatively low ($\sim35\%$). We also explored the effects of varying the mass of the exploding star (and thus the ejecta mass) as well as the pre-supernova inner orbital period. A smaller amount of ejecta mass during the explosion could reduce $w_{12}$ and therefore increase the survival probability. A wider pre-supernova inner orbital period also tends to increase the survival probability by reducing the recoil velocity. However, this advantage diminishes at higher natal kick velocities, where the wider inner binary is more easily disrupted during the supernova explosion. 

The tertiary star is located in a sufficiently wide orbit that no Roche-lobe overflow or other significant binary interaction with the inner binary is expected during its evolution. Consequently, it will evolve nearly as a single star and ultimately form a carbon-oxygen WD with a mass of $\sim 0.52\, M_\odot$. Because the tertiary loses mass during its late evolutionary stages, the outer orbit will expand gradually, resulting in a final orbital period of $\sim 70$\,yr. Combining the evolution of the inner binary and the tertiary star, the system will ultimately evolve into a system consisting of an NS, a He WD, and a carbon-oxygen WD tertiary companion.

\begin{acknowledgments}
The authors were supported by the National Key R\&D Program of China No. 2025YFA1614002, the National Natural Science Foundation of China (NSFC), grant numbers 12588202 and 12041303, the Chinese Academy of Sciences via project JZHKYPT-2021-06. 

This work used data from FAST (https://cstr.cn /31116.02.FAST). FAST is a Chinese national mega-science facility, built and operated by the National Astronomical Observatories, Chinese Academy of Sciences. We all appreciate the excellent performance of FAST and the operation team. 
This publication makes use of data products from the Two Micron All Sky Survey (2MASS), which is a joint project of the University of Massachusetts and the Infrared Processing and Analysis Center/California Institute of Technology, funded by the National Aeronautics and Space Administration and the National Science Foundation. 
The Pan-STARRS1 Surveys (PS1) and the PS1 public science archive have been made possible through contributions by the Institute for Astronomy, the University of Hawaii, the Pan-STARRS Project Office, the Max-Planck Society and its participating institutes, the Max Planck Institute for Astronomy, Heidelberg and the Max Planck Institute for Extraterrestrial Physics, Garching, The Johns Hopkins University, Durham University, the University of Edinburgh, the Queen's University Belfast, the Harvard-Smithsonian Center for Astrophysics, the Las Cumbres Observatory Global Telescope Network Incorporated, the National Central University of Taiwan, the Space Telescope Science Institute, the National Aeronautics and Space Administration under Grant No. NNX08AR22G issued through the Planetary Science Division of the NASA Science Mission Directorate, the National Science Foundation Grant No. AST-1238877, the University of Maryland, Eotvos Lorand University (ELTE), the Los Alamos National Laboratory, and the Gordon and Betty Moore Foundation. 
This work presents results from the European Space Agency (ESA) space mission Gaia. Gaia data are being processed by the Gaia Data Processing and Analysis Consortium (DPAC). Funding for the DPAC is provided by national institutions, in particular the institutions participating in the Gaia Multi-Lateral Agreement (MLA). The Gaia mission website is \url{https://www.cosmos.esa.int/gaia}. The Gaia archive website is \url{https://archives.esac.esa.int/gaia}.
This work used Fermi-LAT data and software obtained from the Fermi Science Support Center (FSSC). 
\end{acknowledgments}

\begin{contribution}
JLH identified problems, directed this project, organized the teamwork, checked the results, and finalized the paper; 
ZLY solved all problems: reprocessed the FAST data, identified the optical tertiary, figured out its properties, found all related formulas caused by the outer orbit, implemented them into {\sc tempo2}, obtained the timing solution from the radio, optical, and Gamma-ray data, and drafted most parts of this paper;
YY processed the Fermi data and got the profiles based on the results from ZLY, and drafted the relevant text; 
BL checked all possible terms of the relativistic effect for this specific triple system, and drafted the relevant text; 
YLG and BW figured out the evolution routine of this triple system, and drafted the relevant text; 
MKY worked out the properties of the sun-like tertiary, and JL, WMG, JNF, and BW joined the relevant discussions; 
LHL worked out the optical plots used in this paper; 
JX maintained the computer platform and data processing systems;
All authors joined the discussions of results and contributed to finalizing the paper. 
\end{contribution}

\facilities{FAST, FERMI, GAIA, 2MASS, PanSTARRS}

\software{
Tempo2 \citep{Hobbs+2006MNRAS.369..655H}, Ultranest \citep{Buchner+2021JOSS....6.3001B}, PSRCHIVE \citep{Hotan+2004PASA...21..302H}, DSPSR \citep{dspsr2011}.
Newly improved functions for the timing model are available on 
https://github.com/Zony-naoc/Timing-model-for-PSR-J0435-3233}

\bibliography{sn-bibliography}{}
\bibliographystyle{aasjournalv7}

\end{document}